\documentclass{aa}
\usepackage[dvips]{graphicx}
\usepackage{psfig}

\newcommand{\lapp}{\mbox{\raisebox{-0.3em}{$\stackrel{\textstyle <}{\sim}$}}}
\newcommand{\gapp}{\mbox{\raisebox{-0.3em}{$\stackrel{\textstyle >}{\sim}$}}}

\begin{document}

\thesaurus{}

\title{CSSs in a sample of B2 radio sources of intermediate strength}

\author{ D.J. Saikia \inst{1,2} \and P. Thomasson \inst{2} \and
R.E. Spencer \inst{2} \and  F. Mantovani \inst{3} \and C.J. Salter \inst{4} 
\and S. Jeyakumar \inst{5} }

\titlerunning{CSSs in a sample of B2 radio sources}

\offprints{D.J. Saikia}

\institute{
National Centre for Radio Astrophysics,
Tata Institute of Fundamental Research, Post Bag No. 3,\\
Ganeshkhind, Pune 411 007, India,
\and
Jodrell Bank Observatory, University of Manchester, Macclesfield, Cheshire SK11 9DL, UK
\and
Istituto di Radioastronomia, CNR, Via P. Gobetti 101, Bologna, Italy
\and
Arecibo Observatory, HC3 Box 53995, Arecibo, Puerto Rico PR 00612, USA 
\and
Physikalisches Institut, Universit\"{a}t zu K\"{o}ln, Z\"{u}lpicher Str. 77, 
50937 K\"{o}ln, Germany
}

\date{Received 0000; Accepted 0000}

\maketitle

\begin{abstract}
We present radio observations of 19 candidate compact steep-spectrum (CSS) objects selected 
from a well-defined, complete sample of 52 B2 radio sources of intermediate strength. 
These observations were made with the VLA A-array at 4.835 GHz.
The radio structures of the entire sample are summarised and the brightness asymmetries
within the compact sources are compared with those of the more extended ones, 
as well as with those in the
3CRR sample and the CSSs from the B3-VLA sample. About 25 per cent of the CSS sources
exhibit large brightness asymmetries, with a flux density ratio for the  opposing
lobes of $>$5, possibly due
to interaction of the jets with infalling material. The corresponding percentage for the
larger-sized objects is only about 5 per cent.  We also investigate possible dependence of 
the  flux density asymmetry of the lobes on redshift, since 
this might be affected by more interactions and mergers in the past. No such 
dependence is found.  A few individual objects of interest are discussed in the paper.
\end{abstract}


\section{Introduction }
Statistical studies of the structures of radio sources have played an 
important role in understanding both their cosmological evolution and
their evolution with age. 
One of the best studied complete samples is the 3CRR 
which has complete identification and redshift information (Laing et al.
1983). Examples of other well-studied samples covering 
significant areas of the sky 
are the Molonglo Reference Catalogue 1-Jy sample (Large et al. 1981; Kapahi
et al. 1998a,b), the 2-Jy all-sky (Wall \& Peacock 1985 and
references therein) and the S4 samples (Pauliny-Toth et al. 1978; Kapahi
1981; Stickel \& K\"{u}hr 1994; Saikia et al. 2001). Studies of compact
steep spectrum sources (CSSs) in such well-defined source samples have
provided useful insights towards understanding the evolution of radio sources
(cf. O'Dea 1998 for a review). In recent years, attention has also been
focussed on CSSs selected from samples that are significantly weaker than 
those investigated earlier (e.g. Fanti et al. 2001; Kunert et al. 2002).

\begin{figure*}[t]
\hbox{
  \vbox{
  \psfig{file=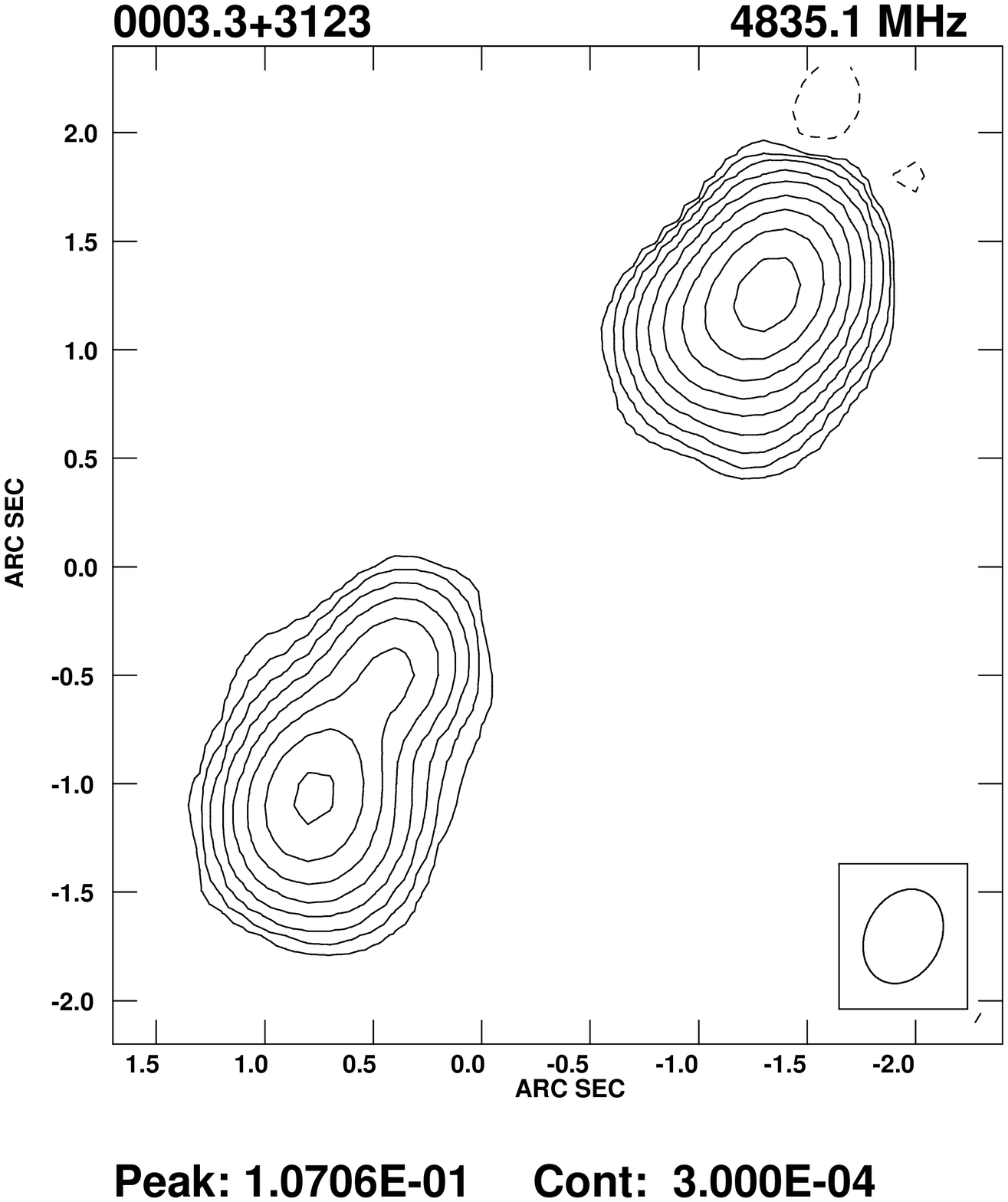,width=2.15in}
  \psfig{file=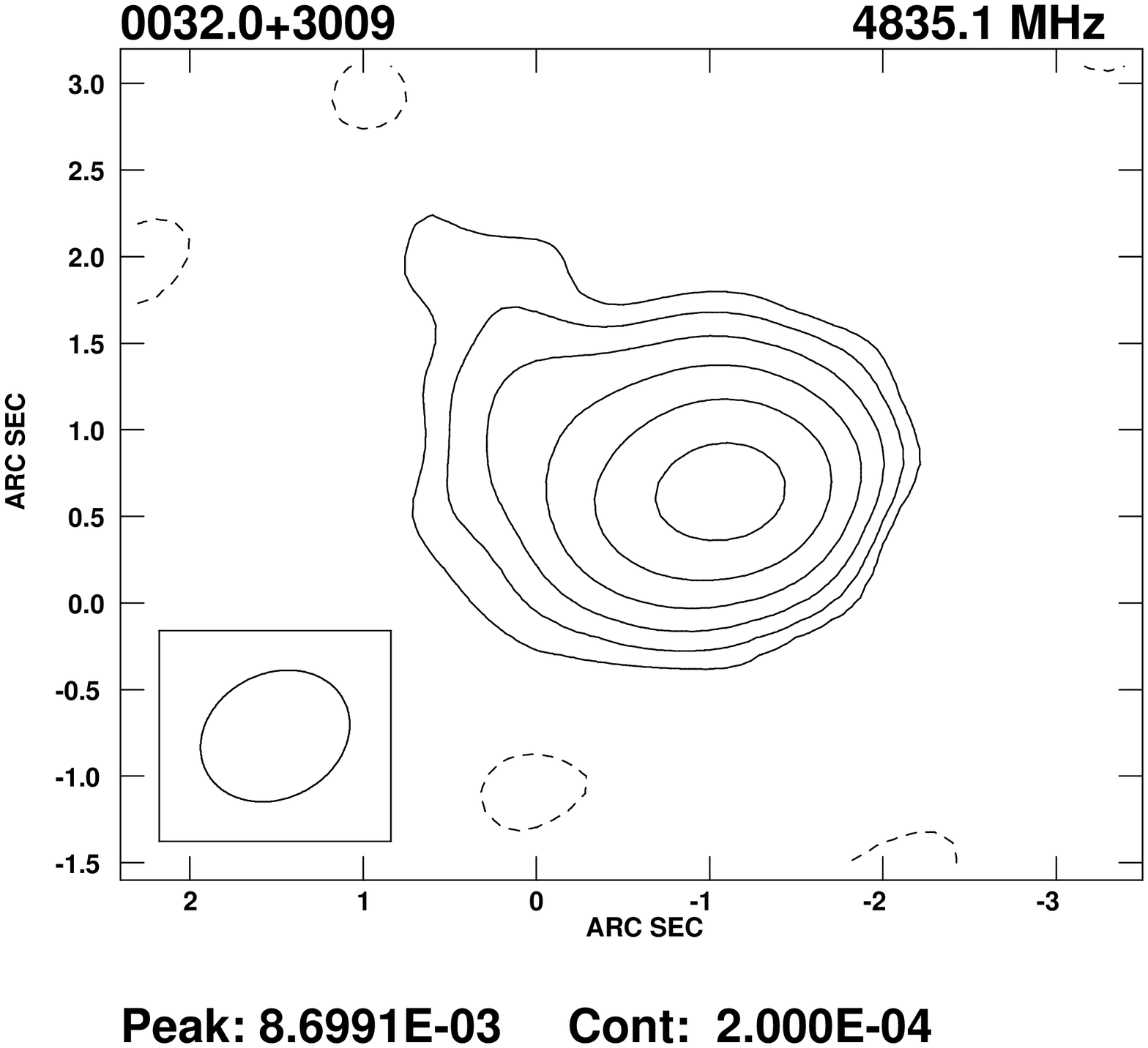,width=2.1in,angle=-90}
  \psfig{file=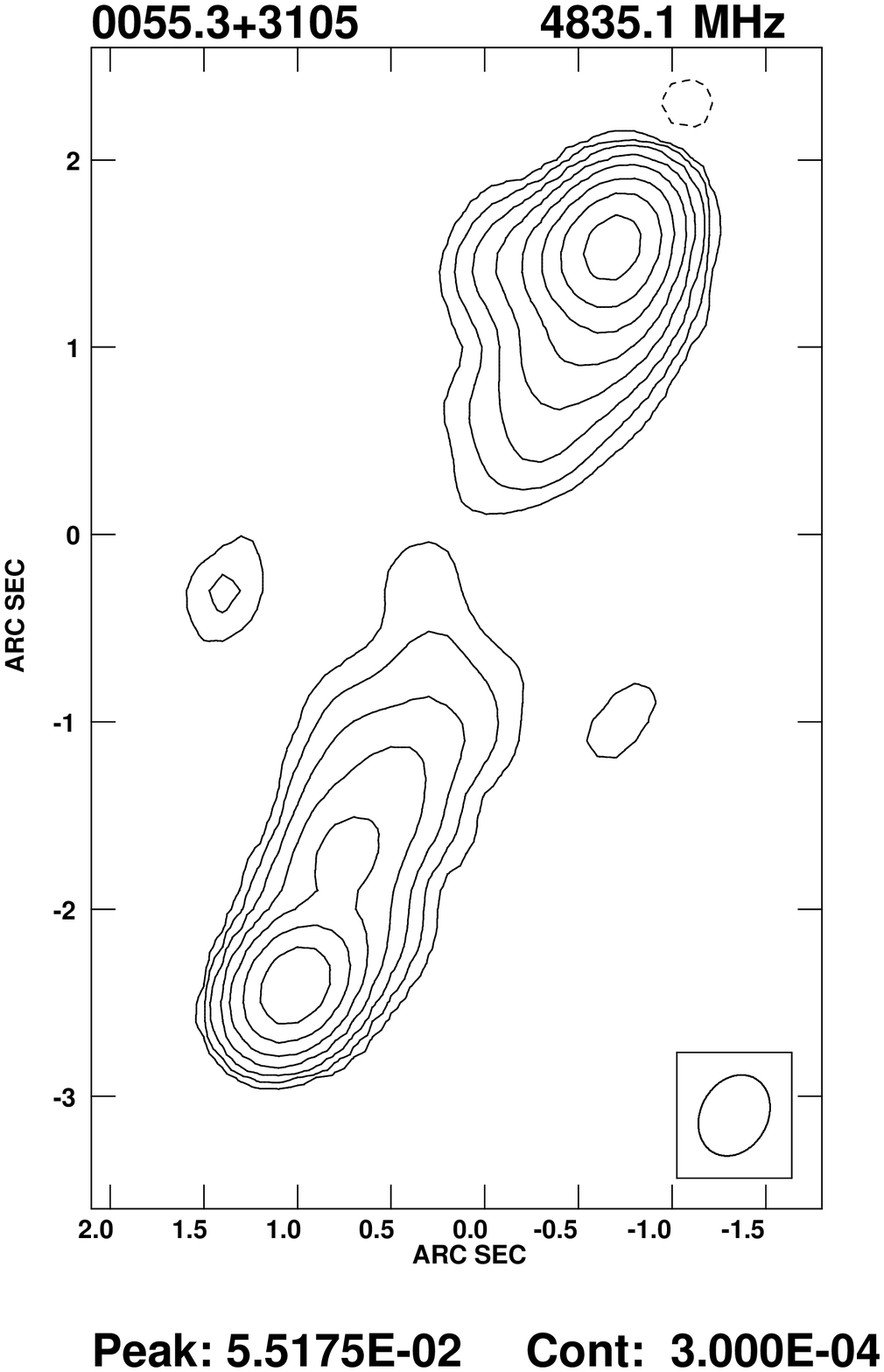,width=1.95in}
  }
  \vbox{
  \psfig{file=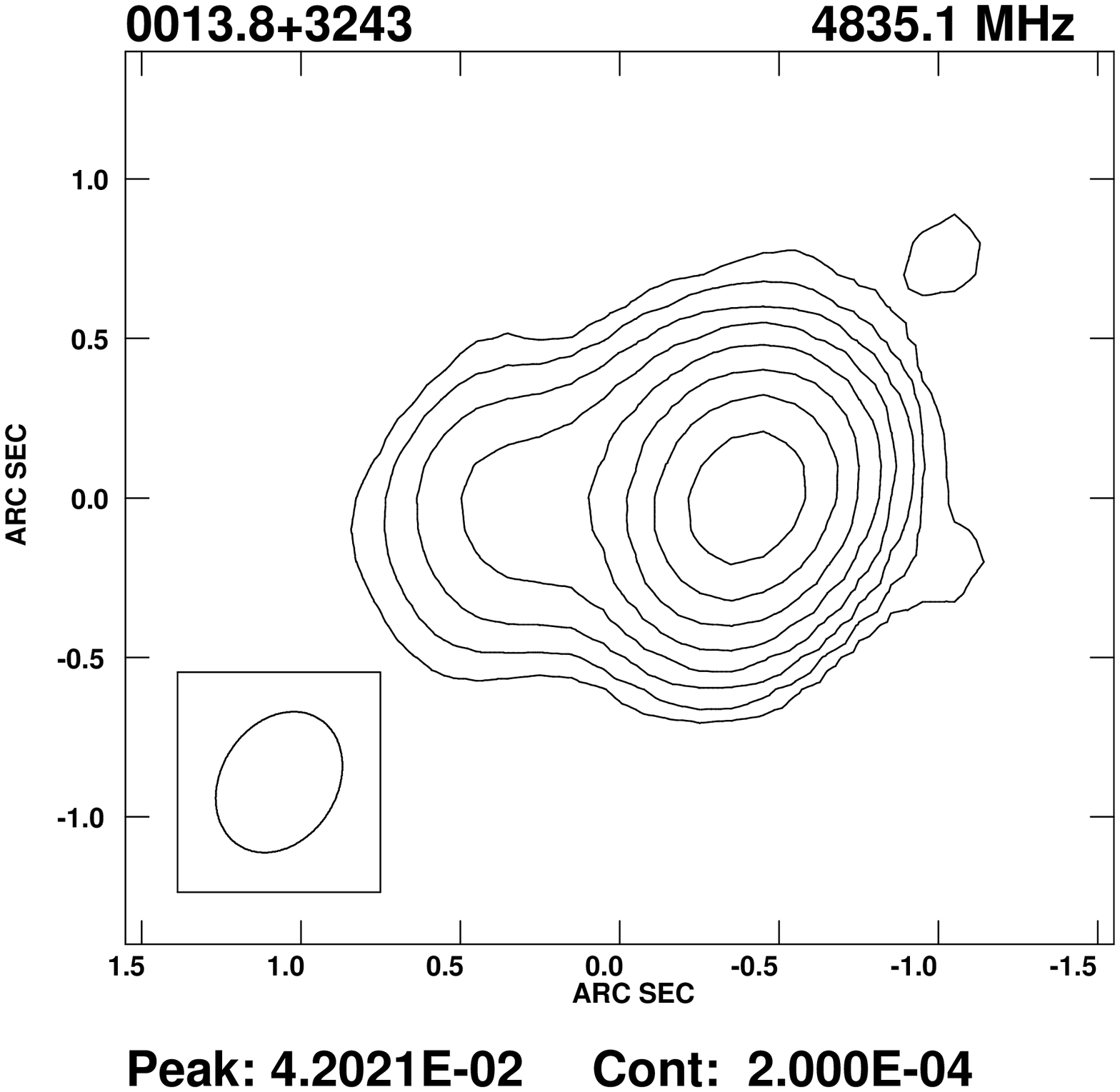,width=2.1in}
  \psfig{file=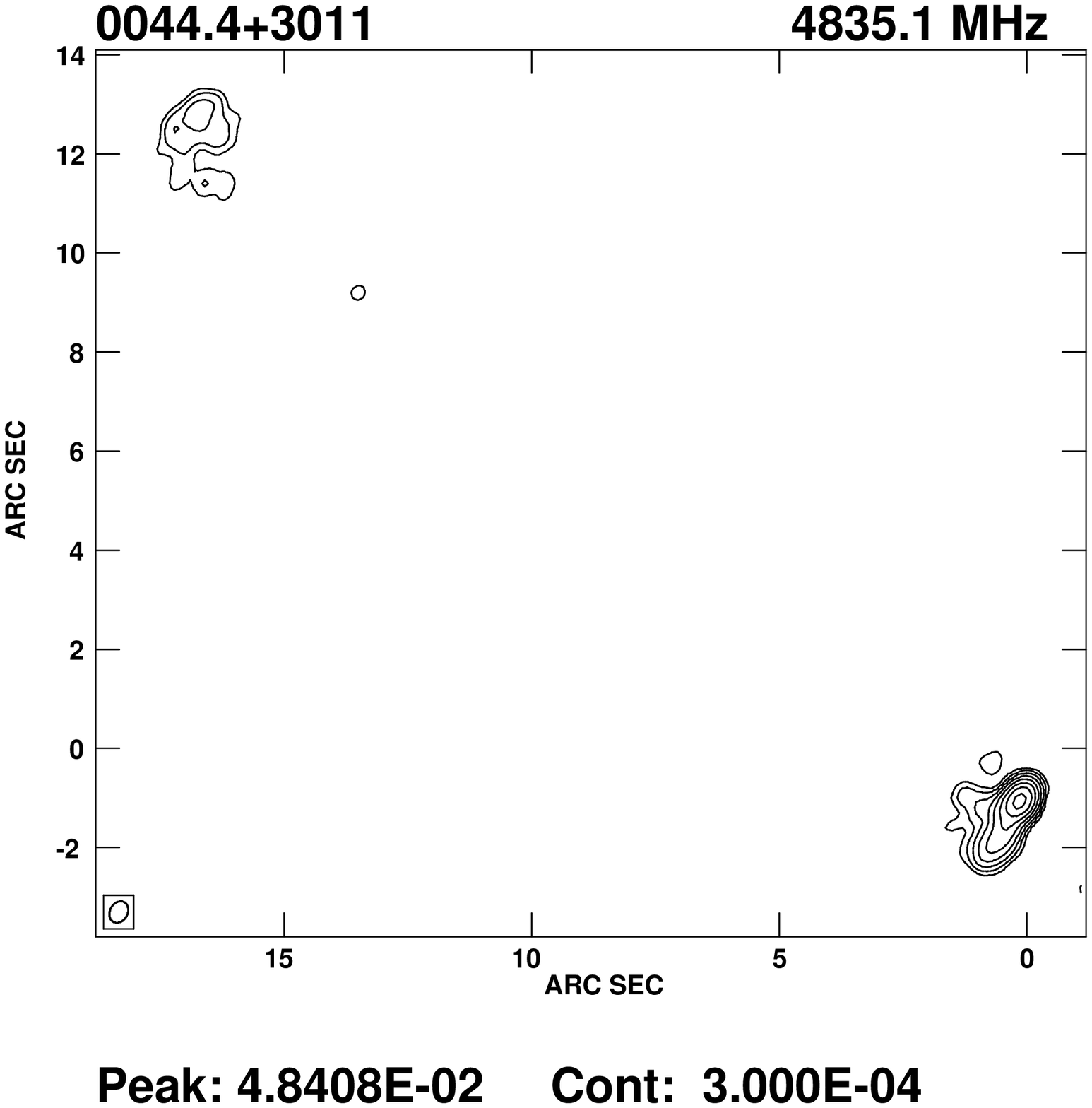,width=2.5in}
  \psfig{file=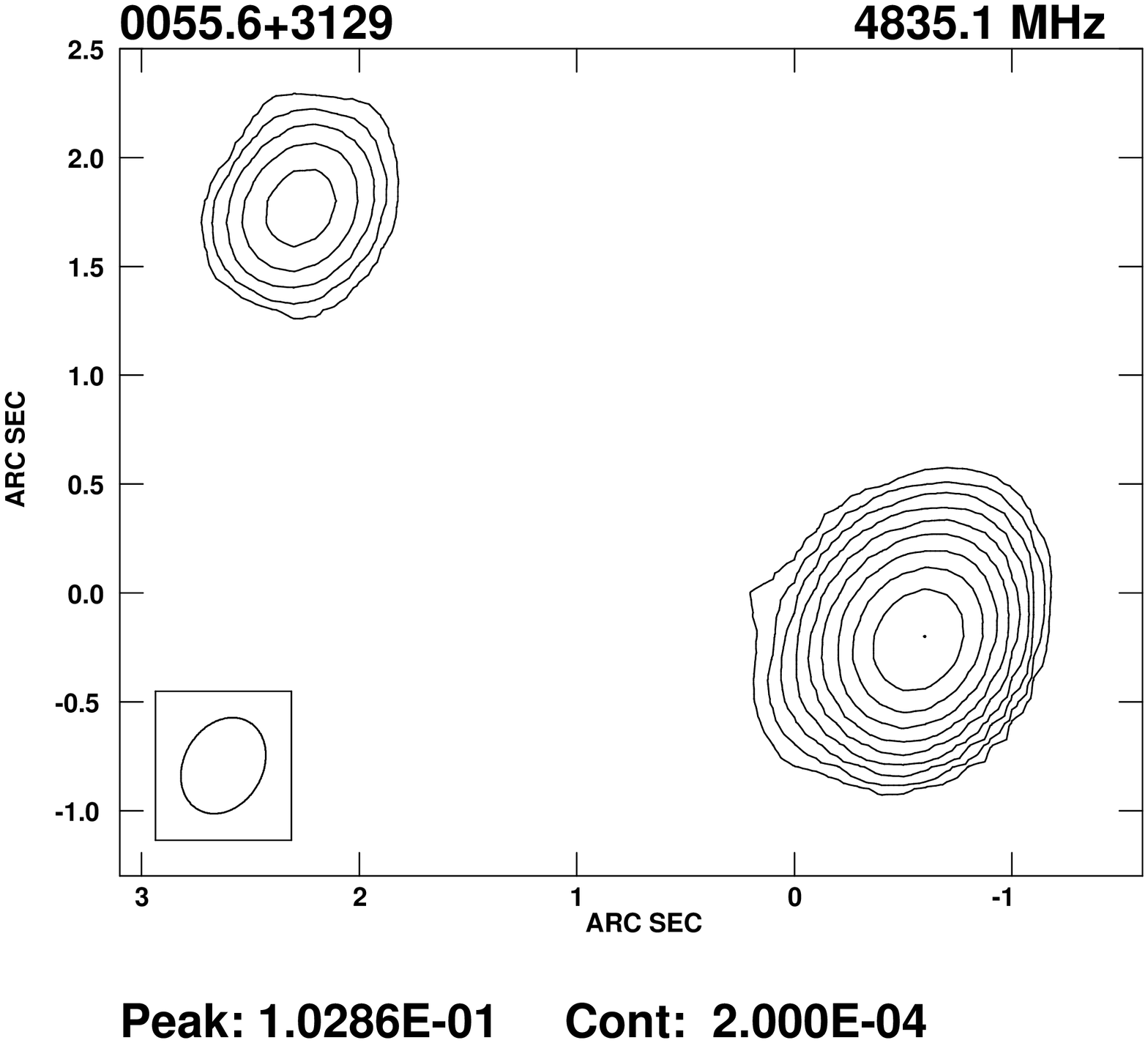,width=2.1in,angle=-90}
  }
  \vbox{
  \psfig{file=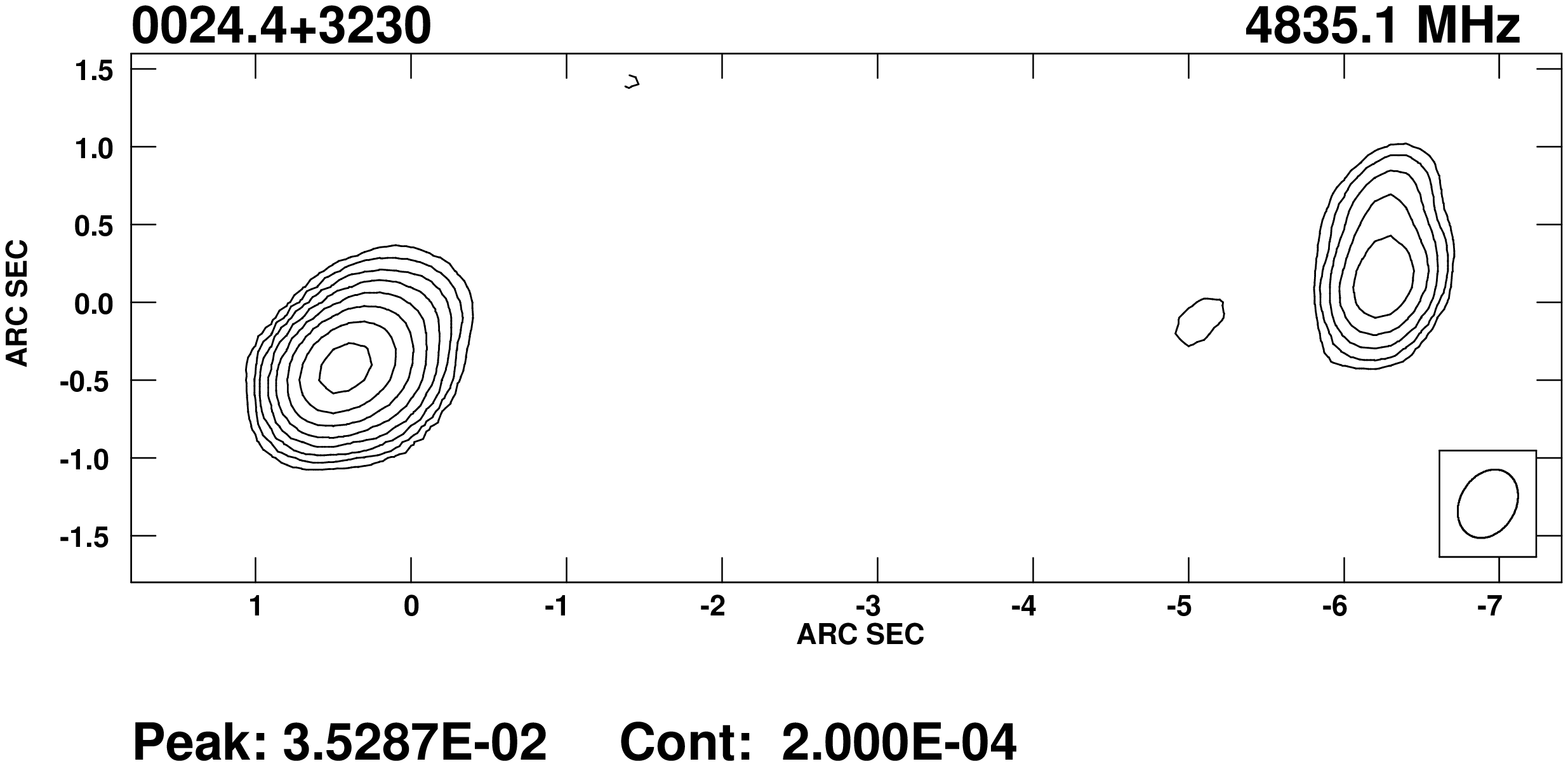,width=2.3in,angle=-90}
  \psfig{file=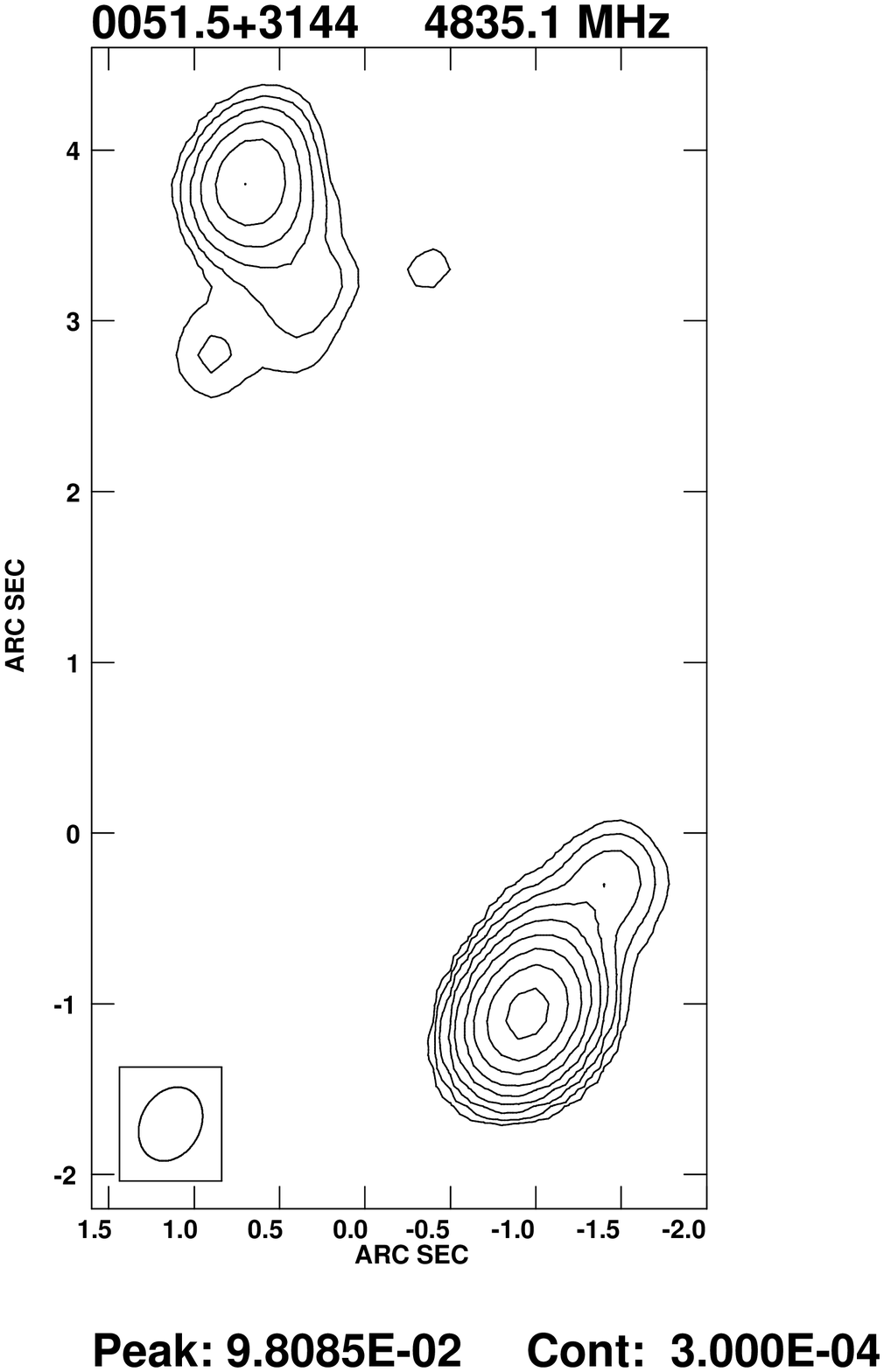,width=1.4in}
  \psfig{file=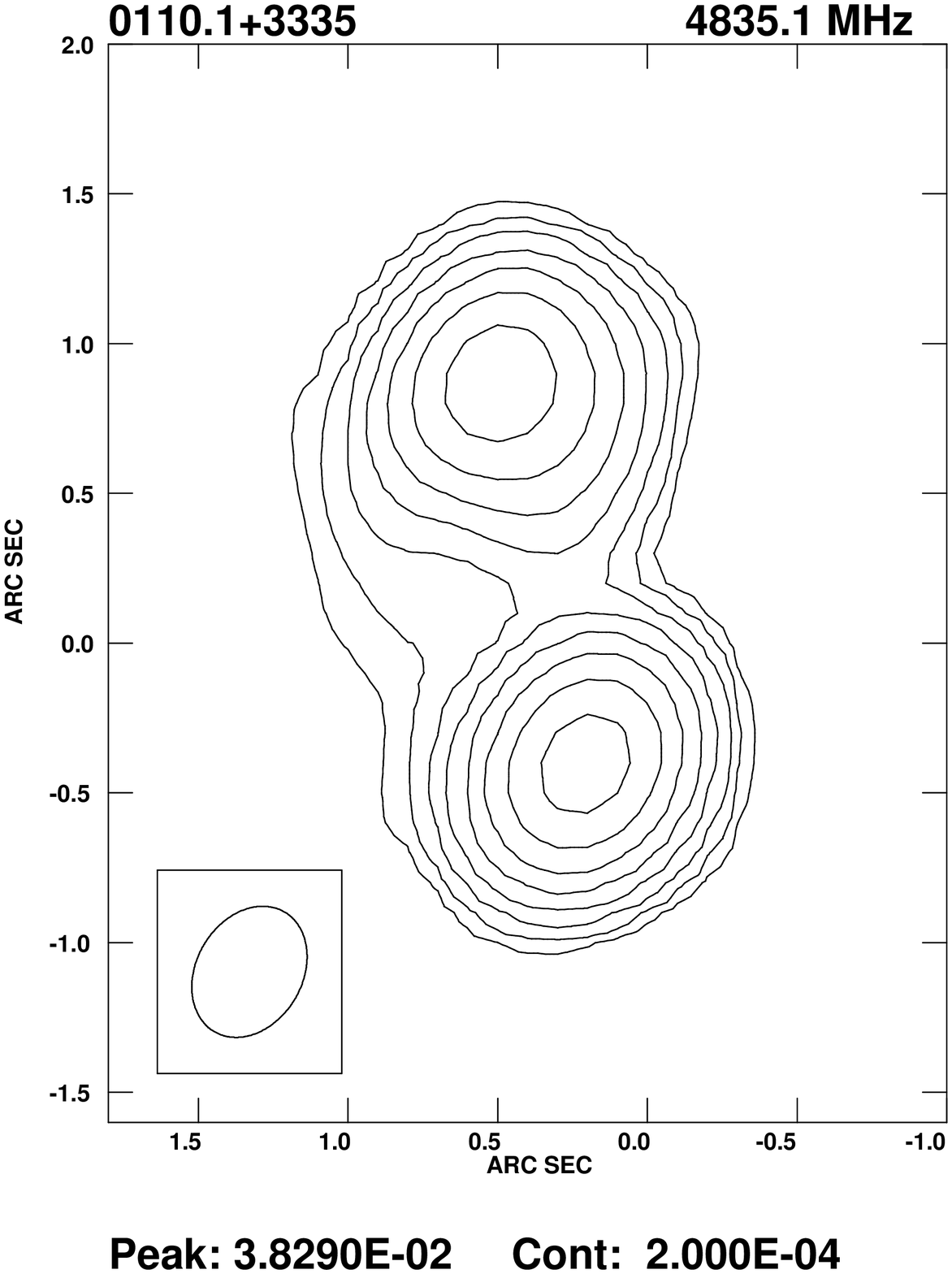,width=1.7in}
  \psfig{file=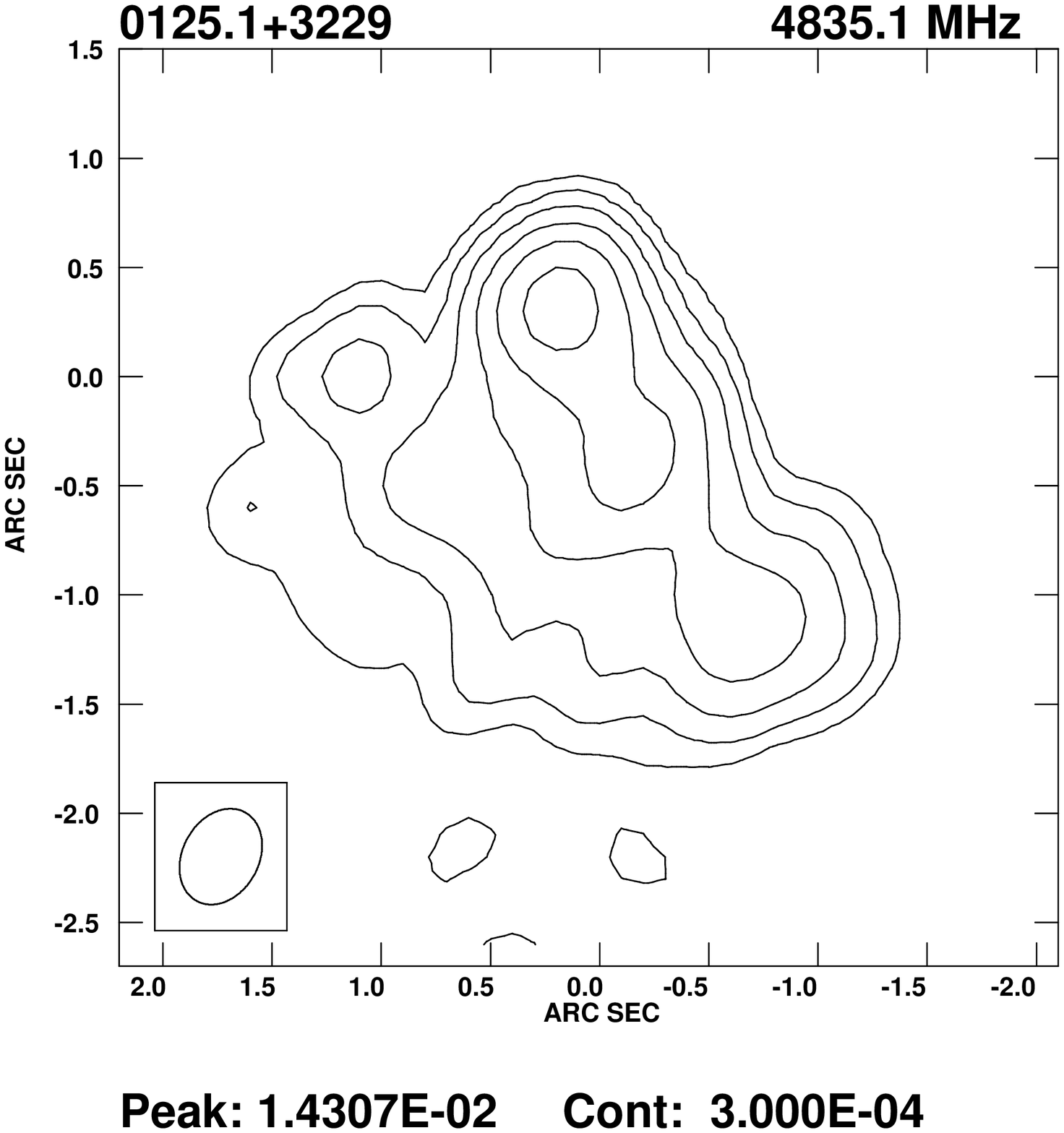,width=1.6in}
  }
}
\caption[]{VLA images of our sample of sources. The contour 
levels for all the images are -1, 1, 2, 4, 8, 16, \ldots times the first contour level. The
peak brightness in the image and the level of the first contour in units of Jy/beam are given 
below the images. The half-power ellipse of the restoring beam is shown in one of the
lower corners of each image.
}
\end{figure*}
 
\begin{figure*}[t]
\hbox{
  \vbox{
  \psfig{file=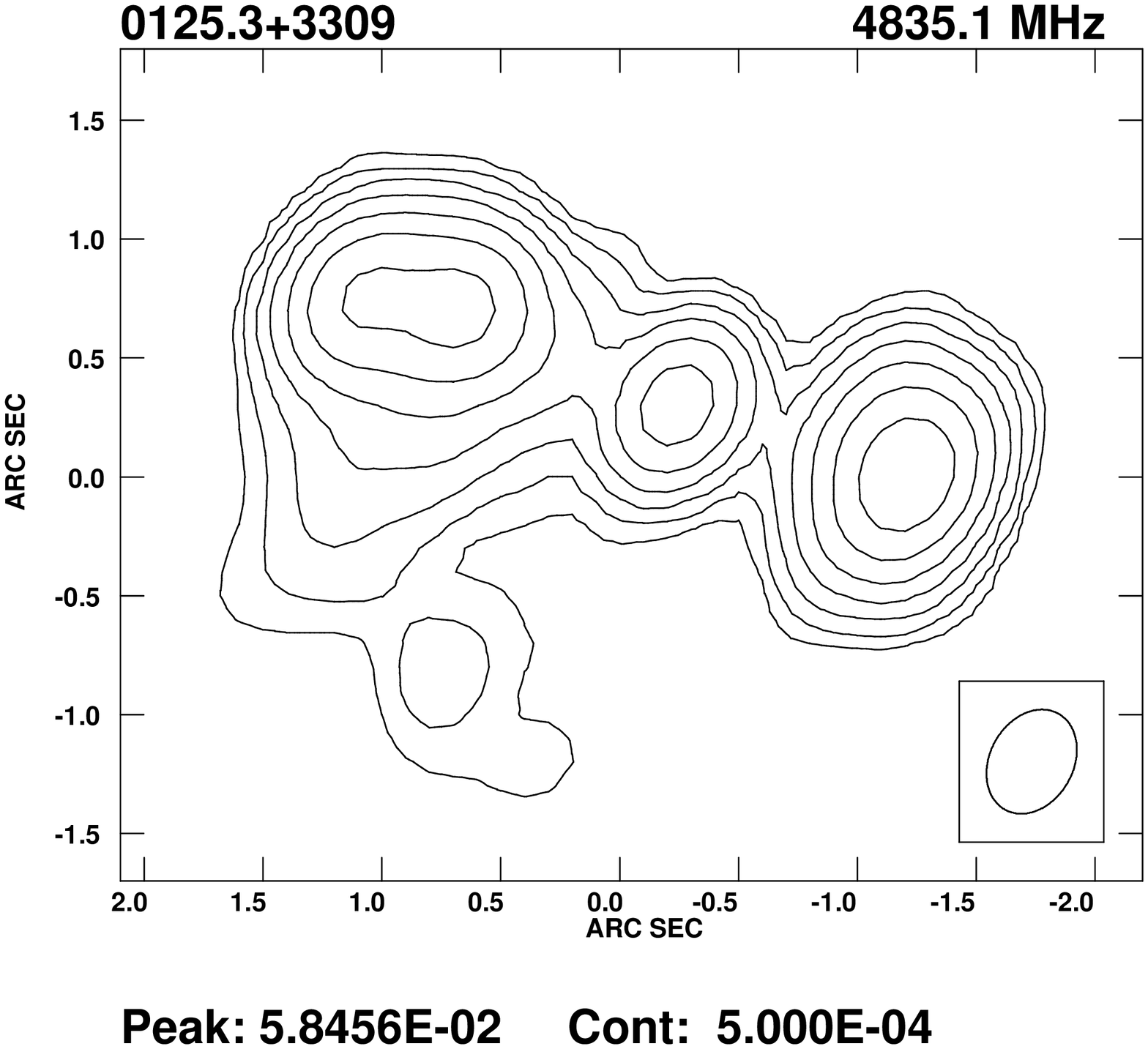,width=2.3in,angle=-90}
  \psfig{file=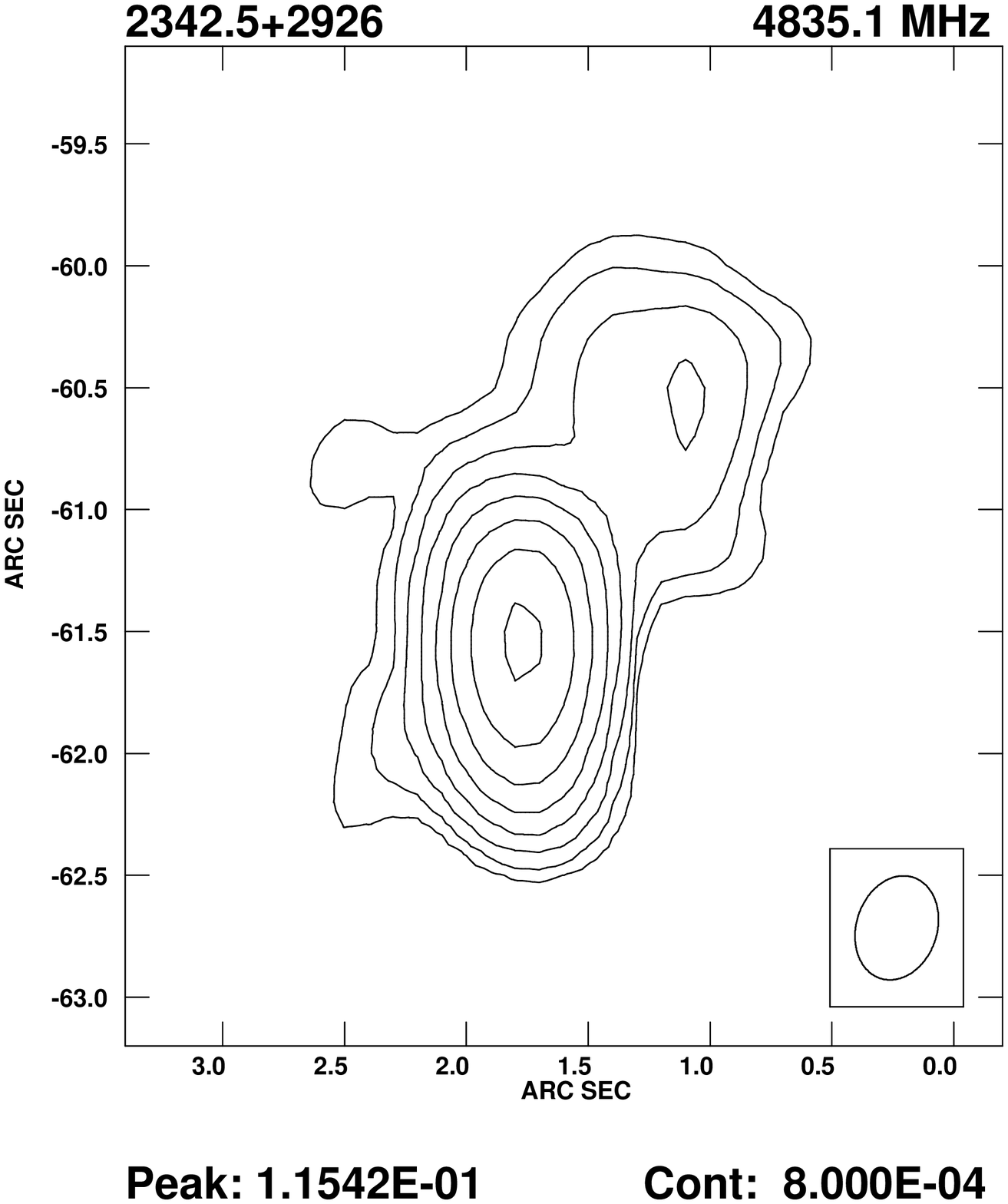,width=1.7in}
  \psfig{file=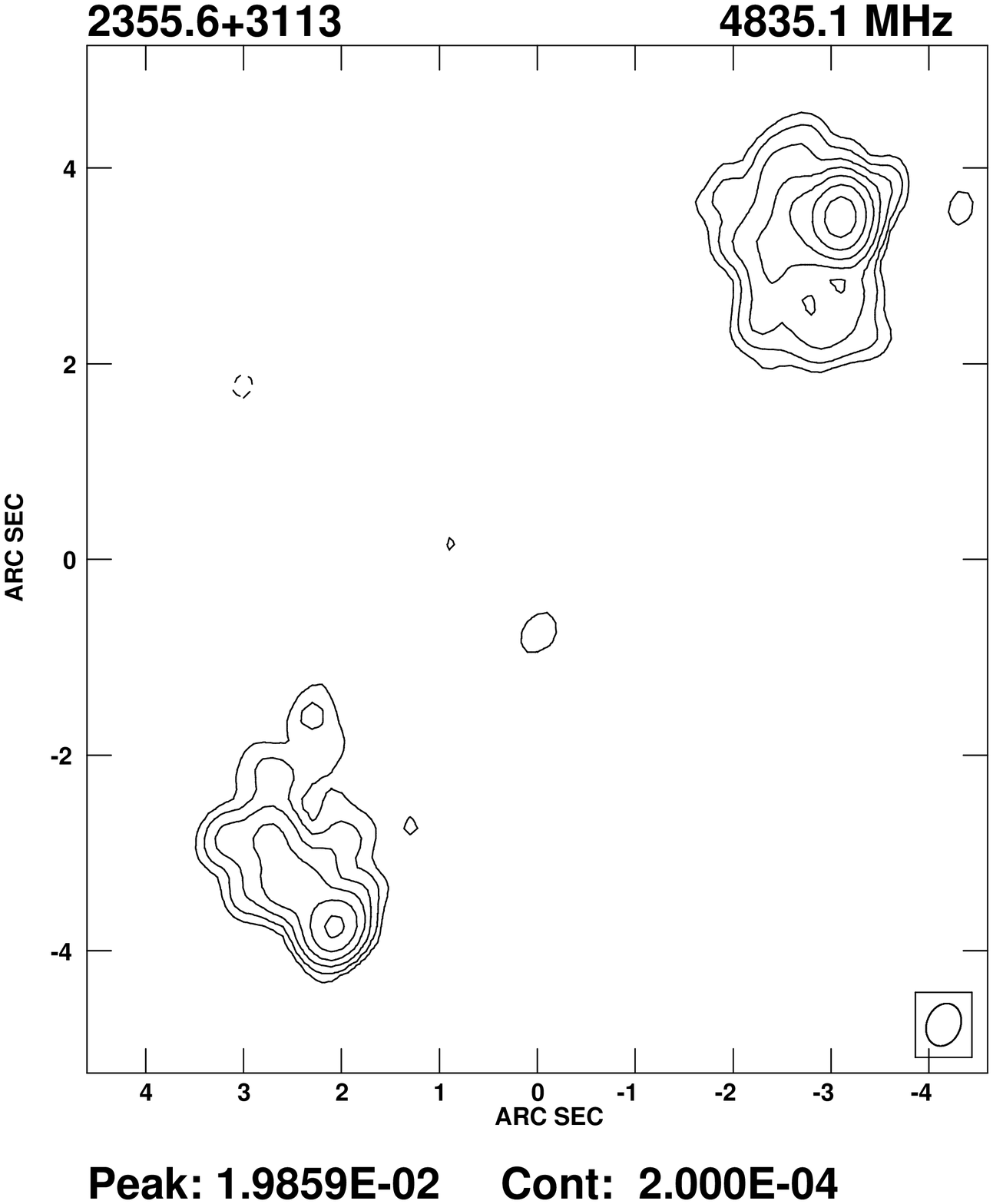,width=1.9in}
  }
  \vbox{
  \psfig{file=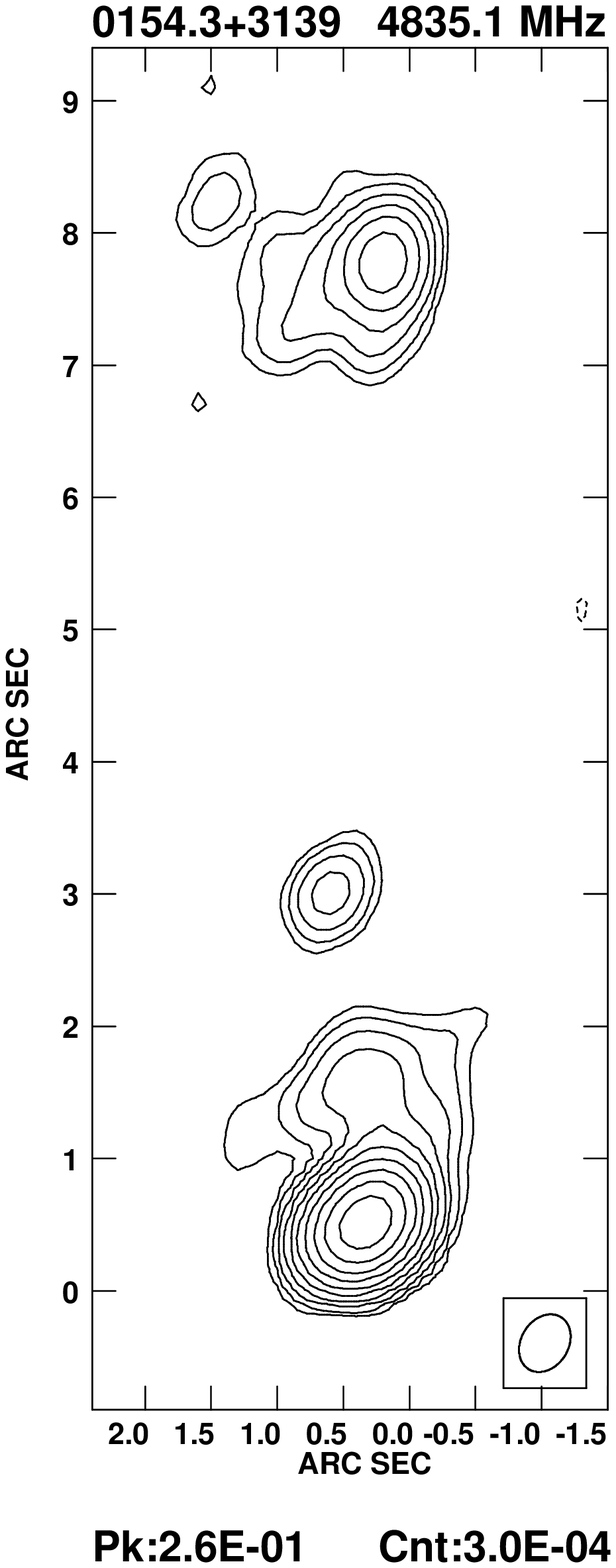,width=1.5in}
  \psfig{file=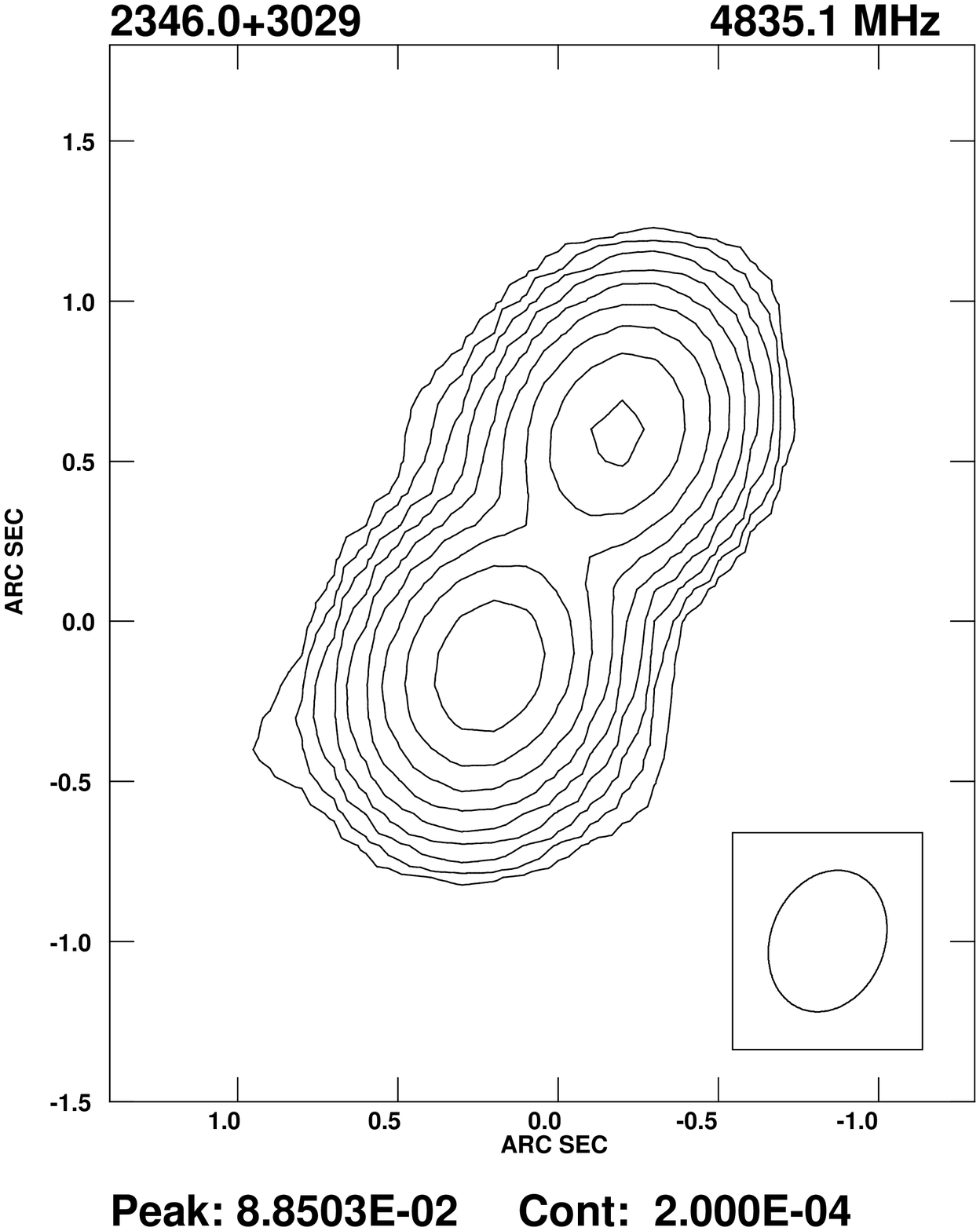,width=1.9in}
  }
  \vbox{
  \psfig{file=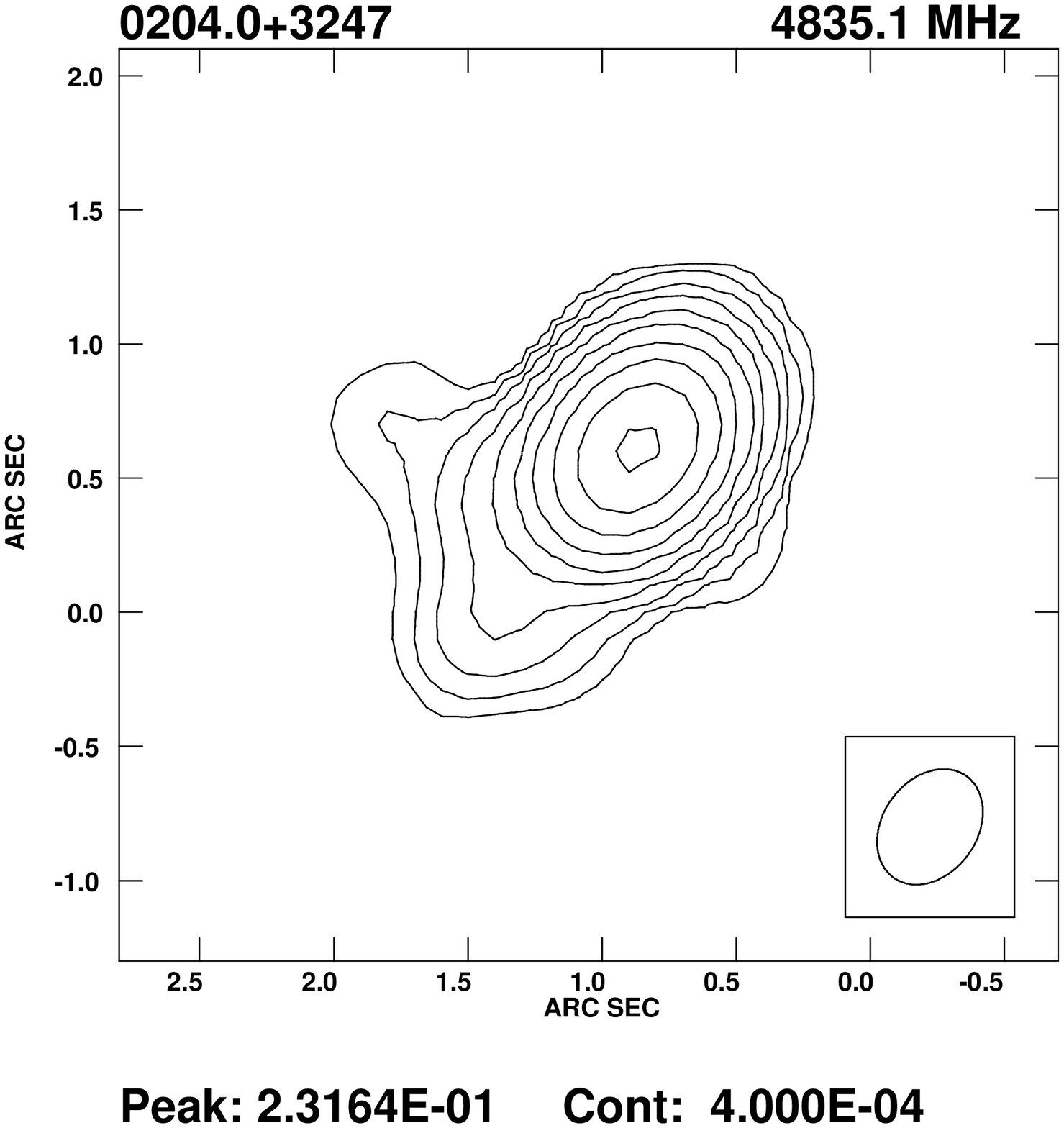,width=1.8in}
  \psfig{file=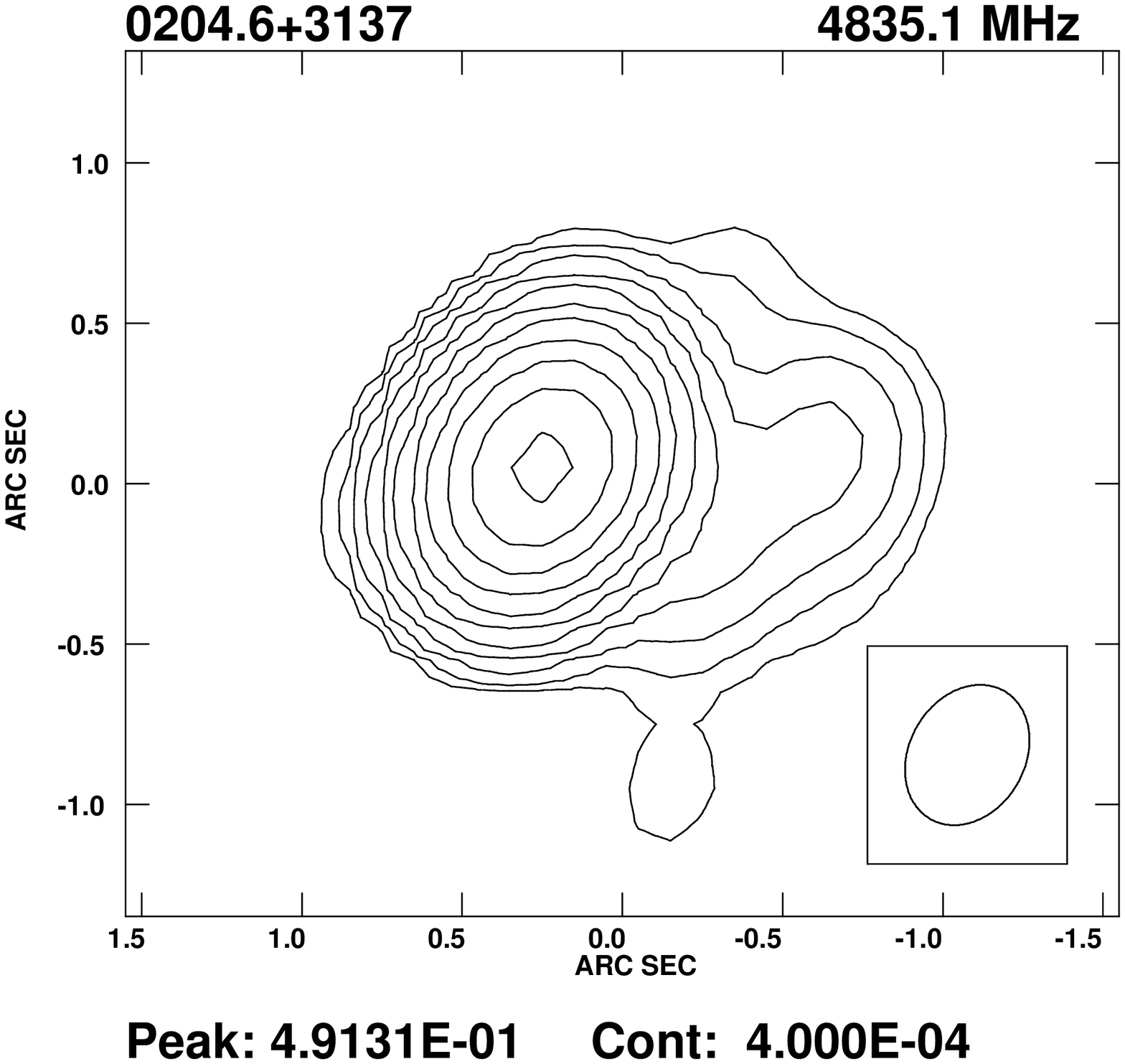,width=1.8in,angle=-90}
  \psfig{file=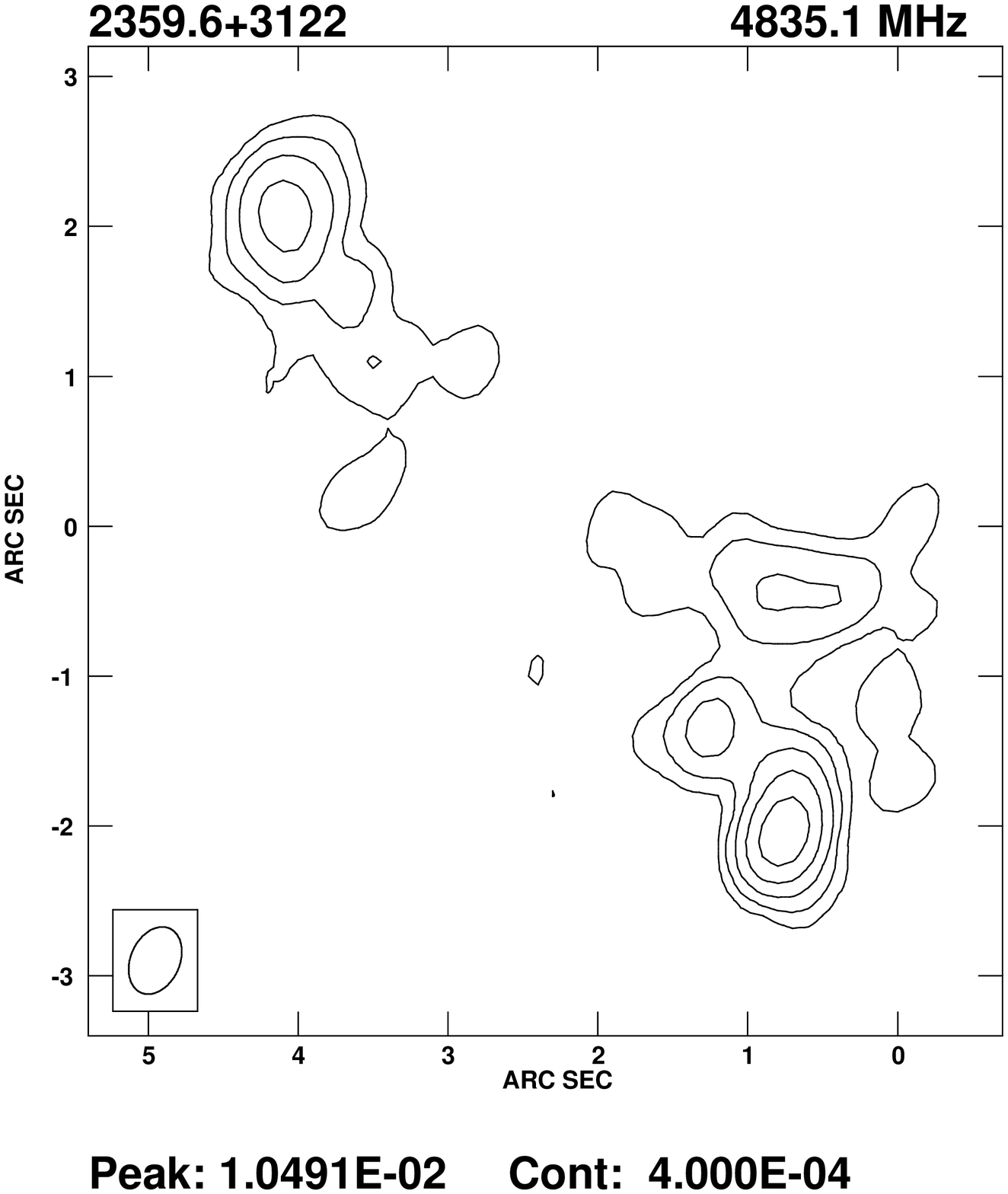,width=1.8in}
  }
}
\caption[]{See the caption to  Fig. 1. 
}
\end{figure*}

The CSSs are physically small objects of subgalactic dimensions. They are
believed to be young sources at an early stage of evolution (Fanti et al. 1995;
Readhead et al. 1996a,b, O'Dea 1998; Owsianik \& Conway 1998). 
Possibly fuelled by the infall of gas, their sizes and structures appear
to be affected by the ambient gas in the central regions of the host galaxies.  
At sub-arcsec resolution, these sources often show strongly distorted structures
with recognizable jet-like features (e.g. Spencer et al. 1991; Schilizzi et al. 2000),
suggesting strong dynamical interaction with the external medium. Evidence 
for the presence of this
gas is also seen in polarization observations of CSSs, which are generally weakly
polarized ($\leq$ 1 per cent) at or below about 5 GHz (Saikia et al. 1985, 1987).
The increase in the median polarization with
frequency (van Breugel et al. 1992; Akujor \& Garrington 1995) suggests
that the depolarization is due to the Faraday effect. A number of CSSs have also
been found to exhibit large rotation measures, leading Mantovani et al. (1994) to 
suggest that most CSSs may be cocooned in dense gaseous envelopes. The radio 
structures of CSSs tend to be more asymmetric than is the
case for the larger sources (Saikia et al.
1995; Jeyakumar \& Saikia 2000; Saikia et al. 2001), indicating an asymmetry in
the distribution of gas on opposite sides of the nucleus on these scales. Evidence
for such asymmetries in the gas distribution, which might be related to the 
fuelling of the radio source, is sometimes provided through a huge difference 
between the rotation measures on opposite sides of the nucleus (Mantovani et al. 1994;
Junor et al. 1999). In addition to the ionized component, there is also evidence
for a high incidence of associated HI absorption towards CSSs (Carilli et al. 1998;
Peck et al. 2000; Morganti et al. 2001).

In this paper, we study the small but well-defined sample of sources of intermediate
strength described in Sect. 2, and concentrate on the radio structures of CSSs 
in this sample.  The radio observations and the results from these 
observations are presented in Sect. 3. The discussion and concluding remarks 
are presented in Sects. 4 and 5 respectively.

%
%

\begin{table*}
{\bf Table 1.} The observational parameters and derived properties for the sources observed \\

\begin{tabular}{l l l rrr r r rr r l r r l}
\hline
Source & Opt. & z & \multicolumn{3}{c}{Beam size} & rms & Cmp. & S$_p$ & S$_{int}$ & S$_t$ & 
Str. & FR & LAS & Notes \\
&  &  & $^{\prime\prime}$ & $^{\prime\prime}$ & $^\circ$ & $\mu$Jy/b & & mJy/b & mJy & mJy &
 & & $^{\prime\prime}$ & \\
\hline
\vspace*{0.3cm}
0003.3+3123 &    &        & 0.45 & 0.35 & 154 &  91 &  N & 107 & 146 & 223 & D  & II & 3.2 & \\
            &    &        &      &      &     &     &  S &  45 &  77 &     &    &    &     &  \\
0013.8+3243 &    &        & 0.47 & 0.36 & 147 &  50 & RS &  42 &  57 &  57 & D? &    & 0.6 &  \\
0024.4+3230 &    &        & 0.47 & 0.36 & 149 &  60 &  W & 5.6 & 8.7 &  61 & D  & II & 6.7 &  \\
            &    &        &      &      &     &     &  E &  35 &  52 &     &    &    &     &  \\
0032.0+3009 & G  & 0.1744 & 0.89 & 0.72 & 117 &  80 & RS & 8.7 &  15 &  15 & SR &    &$\sim$1& \\
0044.4+3011 &    &        & 0.45 & 0.35 & 151 &  64 &  W &  48 &  88 & 101 & D  & II & 22  &  \\
            &    &        &      &      &     &     &  E & 1.9 &  12 &     &    &    &     &  \\
0051.5+3144 &    &        & 0.45 & 0.35 & 152 &  83 &  N & 9.6 &  14 & 137 & D  & II & 5.1 &  \\
            &    &        &      &      &     &     &  S &  98 & 121 &     &    &    &     &  \\
0055.3+3105 &    &        & 0.45 & 0.36 & 151 &  82 &  N &  55 &  82 & 140 & D  & II & 4.3 &  \\
            &    &        &      &      &     &     &  S &  35 &  57 &     &    &    &     &  \\
0055.6+3129 & G  &        & 0.47 & 0.36 & 150 &  67 &  W & 103 & 119 & 125 & D  & II & 3.5 &  \\
            &    &        &      &      &     &     &  E & 4.8 & 5.5 &     &    &    &     &  \\
0110.1+3335 &    &        & 0.46 & 0.35 & 149 &  62 &  N &  21 &  29 &  72 & D  & II & 1.3 &  \\
            &    &        &      &      &     &     &  S &  38 &  43 &     &    &    &     &  \\
0125.1+3229 &    &        & 0.46 & 0.35 & 151 &  84 & RS &  14 &  64 &  64 &  ? &    & 2.1 & n\\
0125.3+3309 &    &        & 0.46 & 0.35 & 151 & 148 &  W &  58 &  86 & 234 &  T & II & 2.3 &  \\
            &    &        &      &      &     &     &  C &  24 &  28 &     &    &    &     &  \\
            &    &        &      &      &     &     &  E &  45 & 119 &     &    &    &     &  \\
0154.3+3139 & Q  & 0.373  & 0.46 & 0.36 & 148 &  84 &  N &  16 &  30 & 371 &  T & II & 7.3 &  \\
            &    &        &      &      &     &     &  C & 3.6 & 3.6 &     &    &    &     &  \\
            &    &        &      &      &     &     &  S & 260 & 338 &     &    &    &     &  \\ 
0204.0+3247 &    &        & 0.47 & 0.35 & 144 &  69 & RS & 232 & 260 & 260 & D? &    & 0.8 &  \\
0204.6+3137 & G  & 1.213  & 0.46 & 0.36 & 150 & 158 & RS & 491 & 507 & 503 & CJ?&    & 0.9 &n \\
2330.5+2937 &    &        & 0.46 & 0.35 & 160 &  40 & RS &  81 &  82 &  82 &  U &   &$<$0.1&n \\
2342.5+2926 & G  & 0.13069& 0.44 & 0.33 & 162 & 104 &  N & 7.0 &  13 & 263 &  D & II?& 1.2 &n \\
            &    &        &      &      &     &     &  S & 115 & 239 &     &    &    &     &  \\
2346.0+3029 &    &        & 0.45 & 0.35 & 158 &  50 &  N &  60 &  64 & 160 &  D & II & 0.8 &  \\
            &    &        &      &      &     &     &  S &  89 &  91 &     &    &    &     &  \\
2355.6+3113 &    &        & 0.45 & 0.34 & 158 &  48 &  N &  20 &  46 &  70 &  D & II & 8.9 &  \\
            &    &        &      &      &     &     &  S & 7.6 &  23 &     &    &    &     &  \\ 
2359.6+3122 &    &        & 0.47 & 0.33 & 156 &  86 &  W &  10 &  35 &  62 &  D & II & 5.3 &  \\
            &    &        &      &      &     &     &  E & 4.5 &  18 &     &    &    &     &  \\
\hline 
\end{tabular}
\end{table*}


\section{The B2 intermediate-strength sample}
The sample of sources of intermediate strength selected from the 408-MHz Bologna 
B2.1 catalogue (Colla et al. 1970) consists of 52 objects 
which satisfy the following criteria.
They lie within the region, 23$^h$ 30$^m$ $<$ RA(B1950) $<$ 2$^h$ 30$^m$
and 29$^\circ$ 18$^\prime$ $<$ Dec(B1950) $<$ 34$^\circ$ 02$^\prime$,
and their flux densities are in the range 0.9$\leq$S$_{408}$$<$2.5 Jy.
These sources have been observed earlier with the Mk IA $-$ Defford
interferometer at 408 MHz (Warwick 1977), and the Westerbork Synthesis
Radio Telescope (WSRT) at 5 GHz (Padrielli et al. 1981, hereinafter referred
to as P81). Optical identifications and
redshift measurements are not available for most of the sources at present, although
the predicted median redshift for a 408-MHz 1-Jy sample for a range of models
is in very close agreement and yields a value of about 1.2 (cf. Dunlop \& Peacock 1990).

The radio structures of many of the sources in this sample, especially the 
smaller ones, are not
clear from the earlier observations. The main objective of the present project
is to investigate the structure and properties of 
CSSs of intermediate strength, and compare these with the larger sources using 
a well-defined source sample selected at a low frequency. 

\begin{figure*}
\vbox{
\hbox{
  \psfig{file=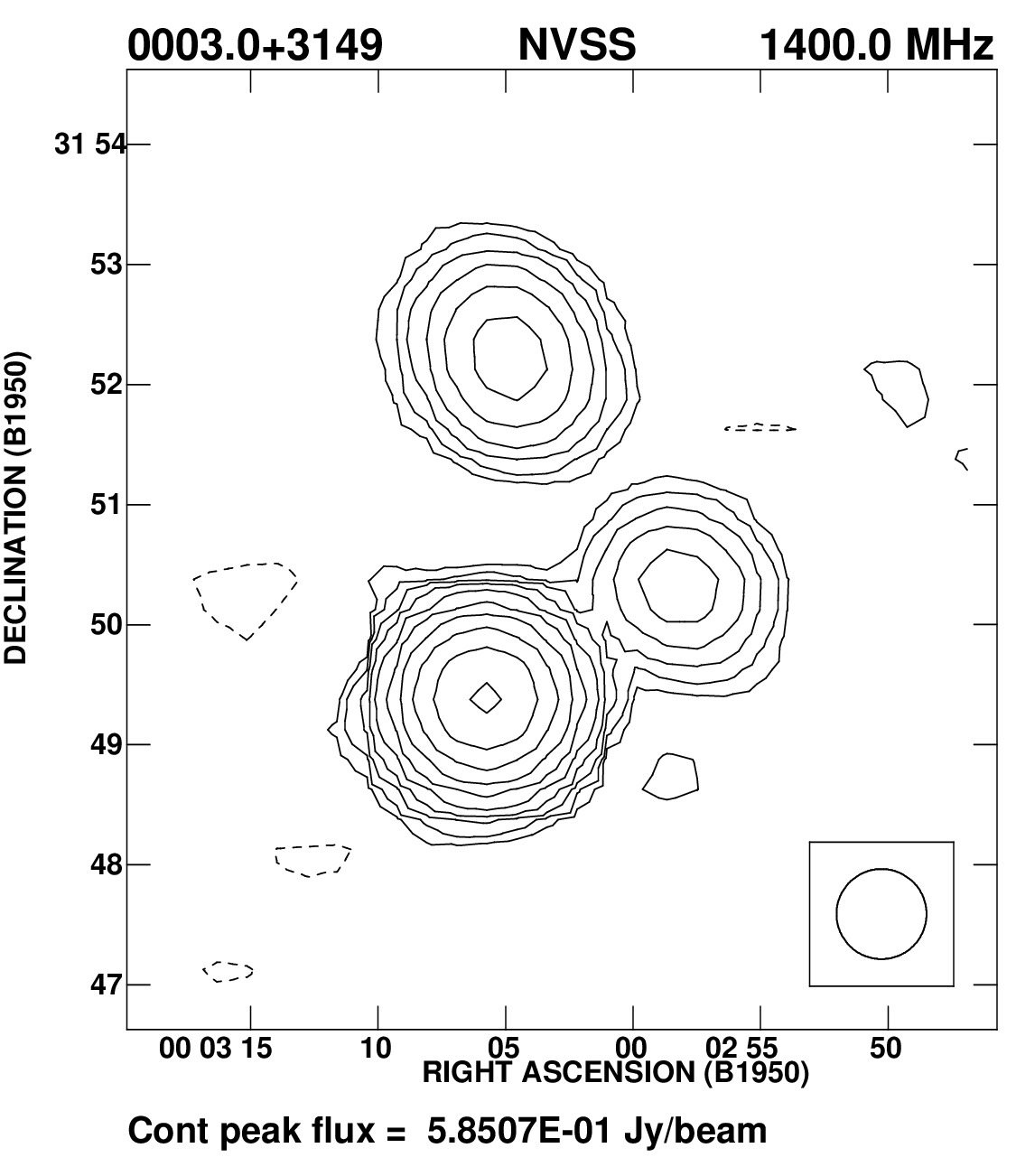,width=1.6in}
  \psfig{file=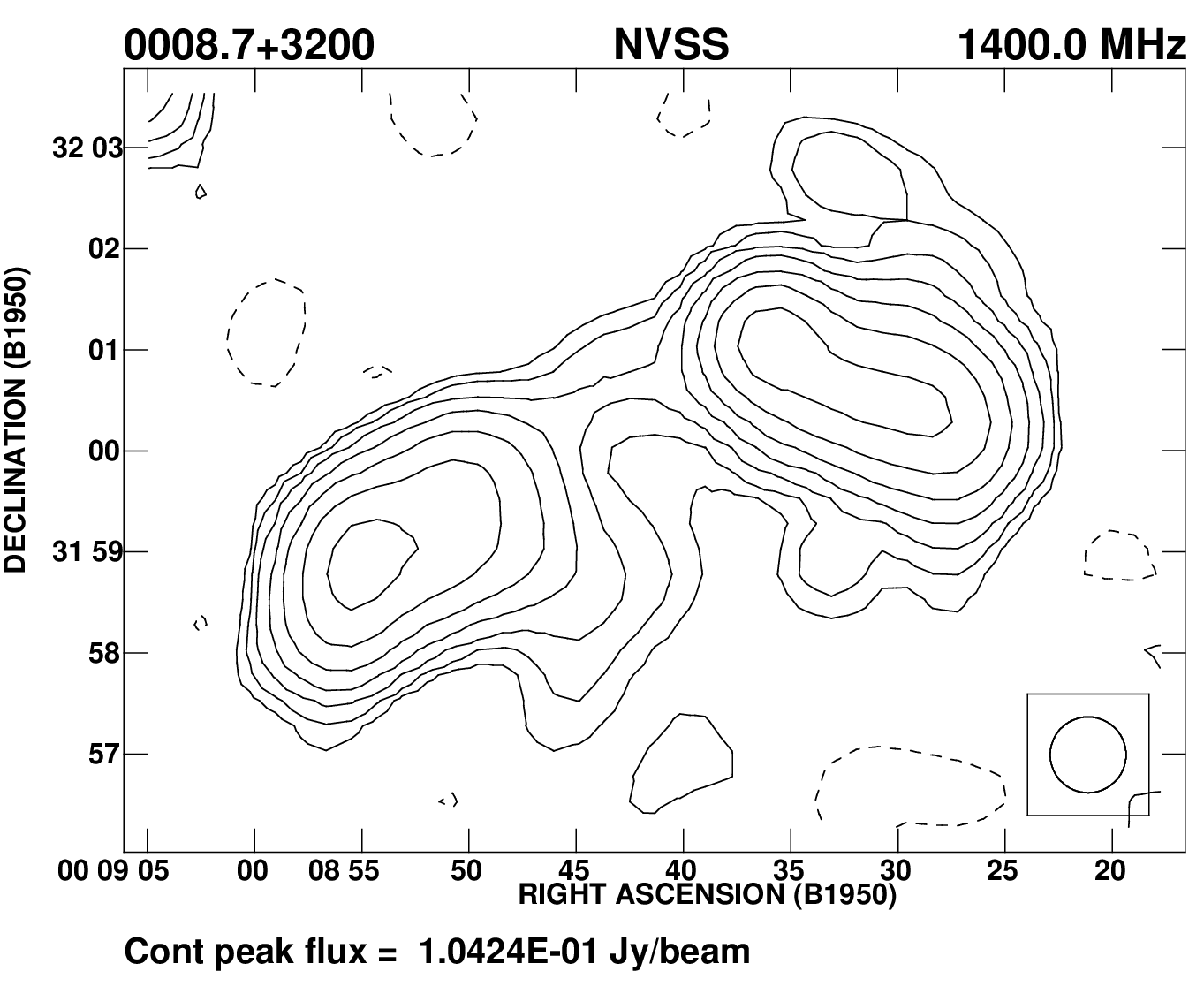,width=2.0in,angle=-90}
  \psfig{file=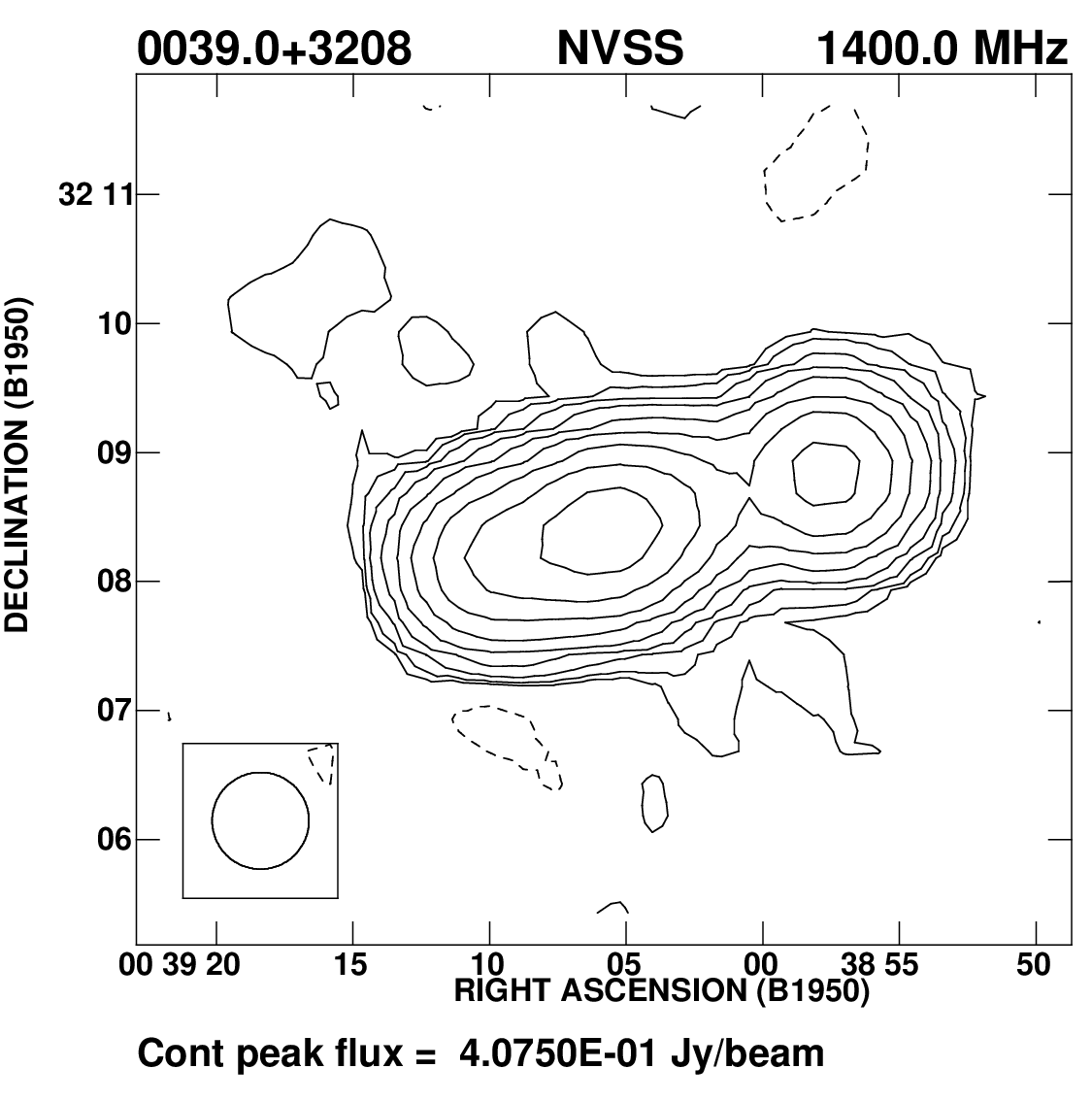,width=1.75in}
  \psfig{file=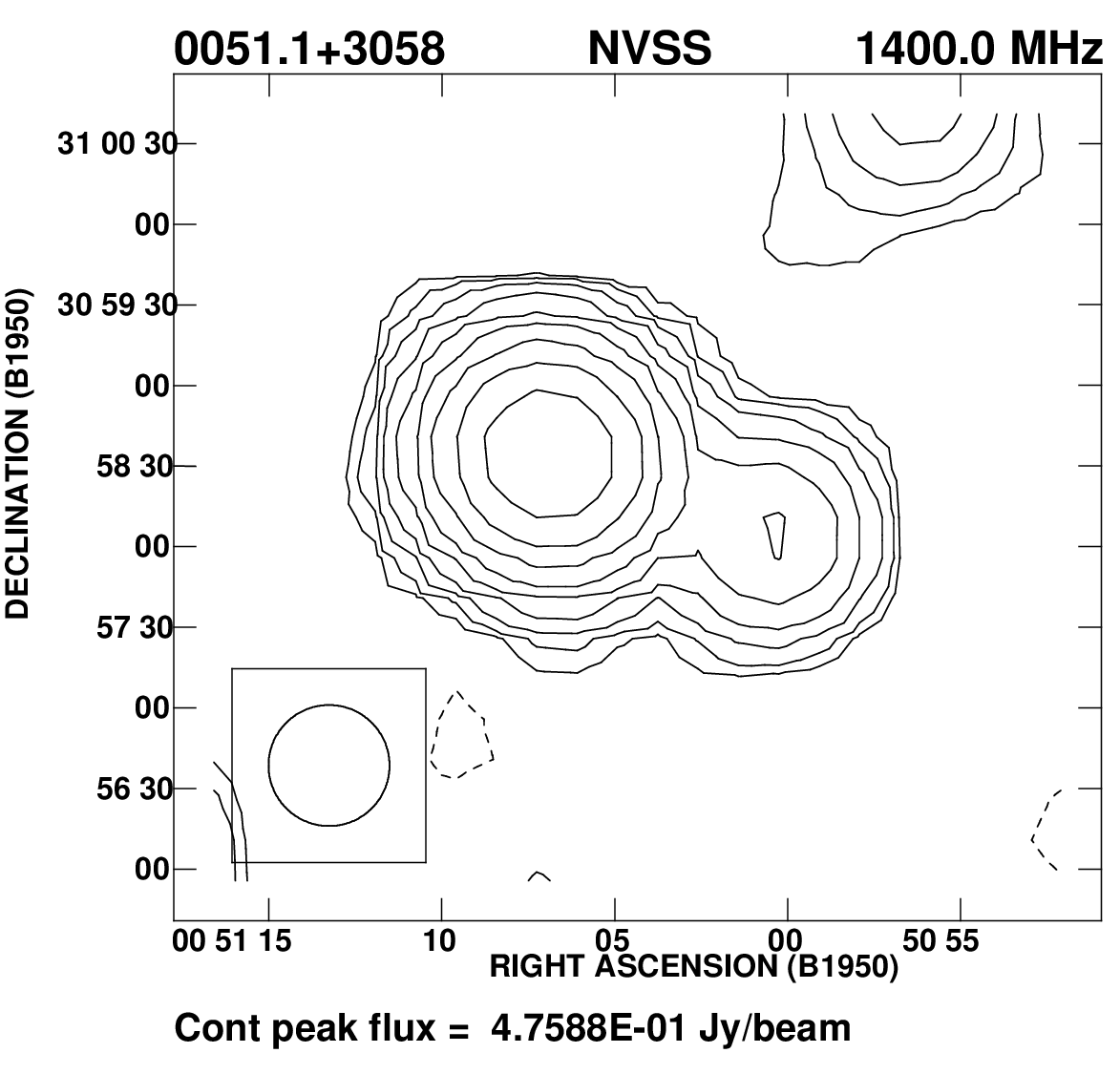,width=1.6in,angle=-90}
     }
\hbox{
  \psfig{file=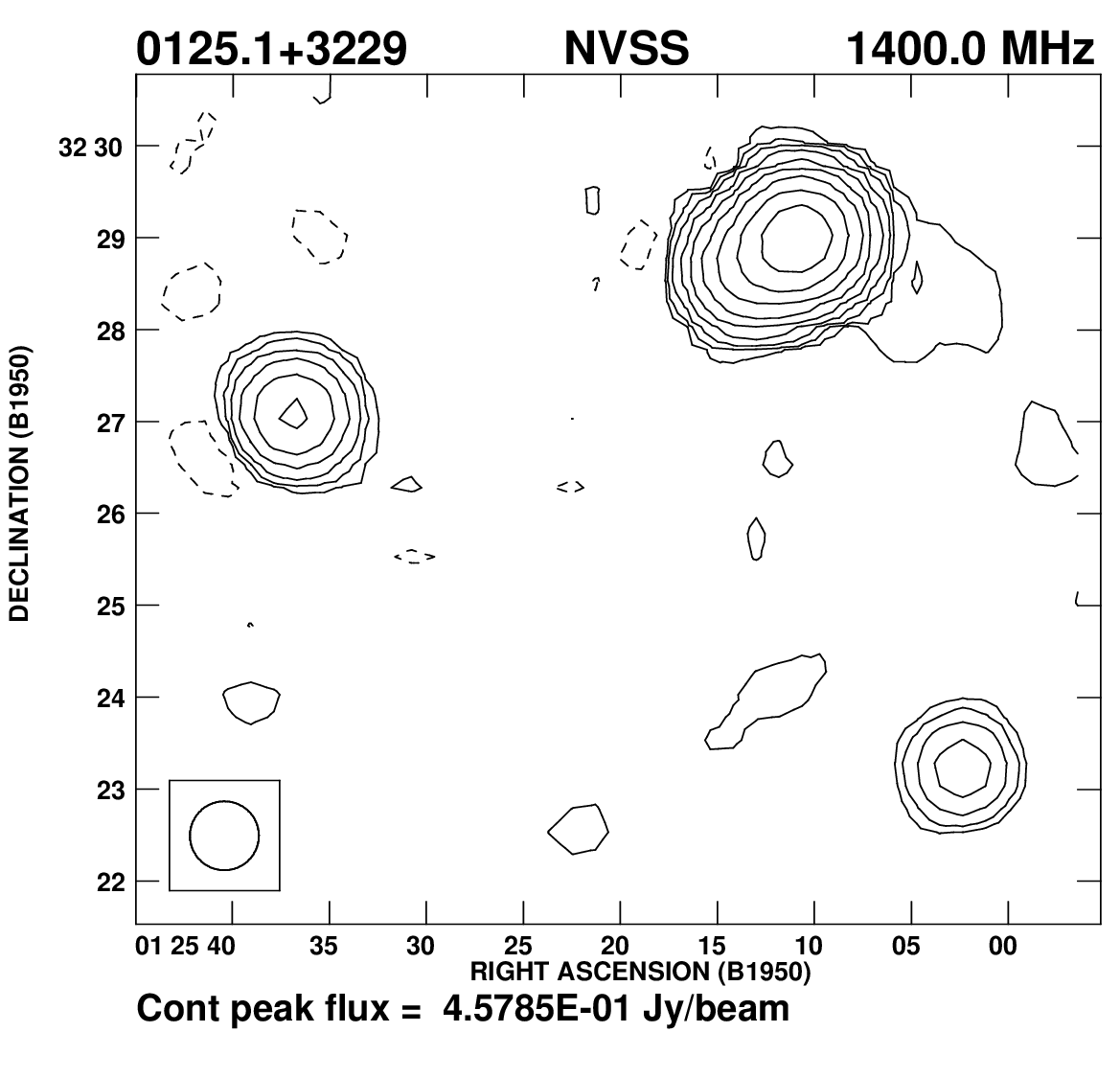,width=1.72in,angle=-90}
  \psfig{file=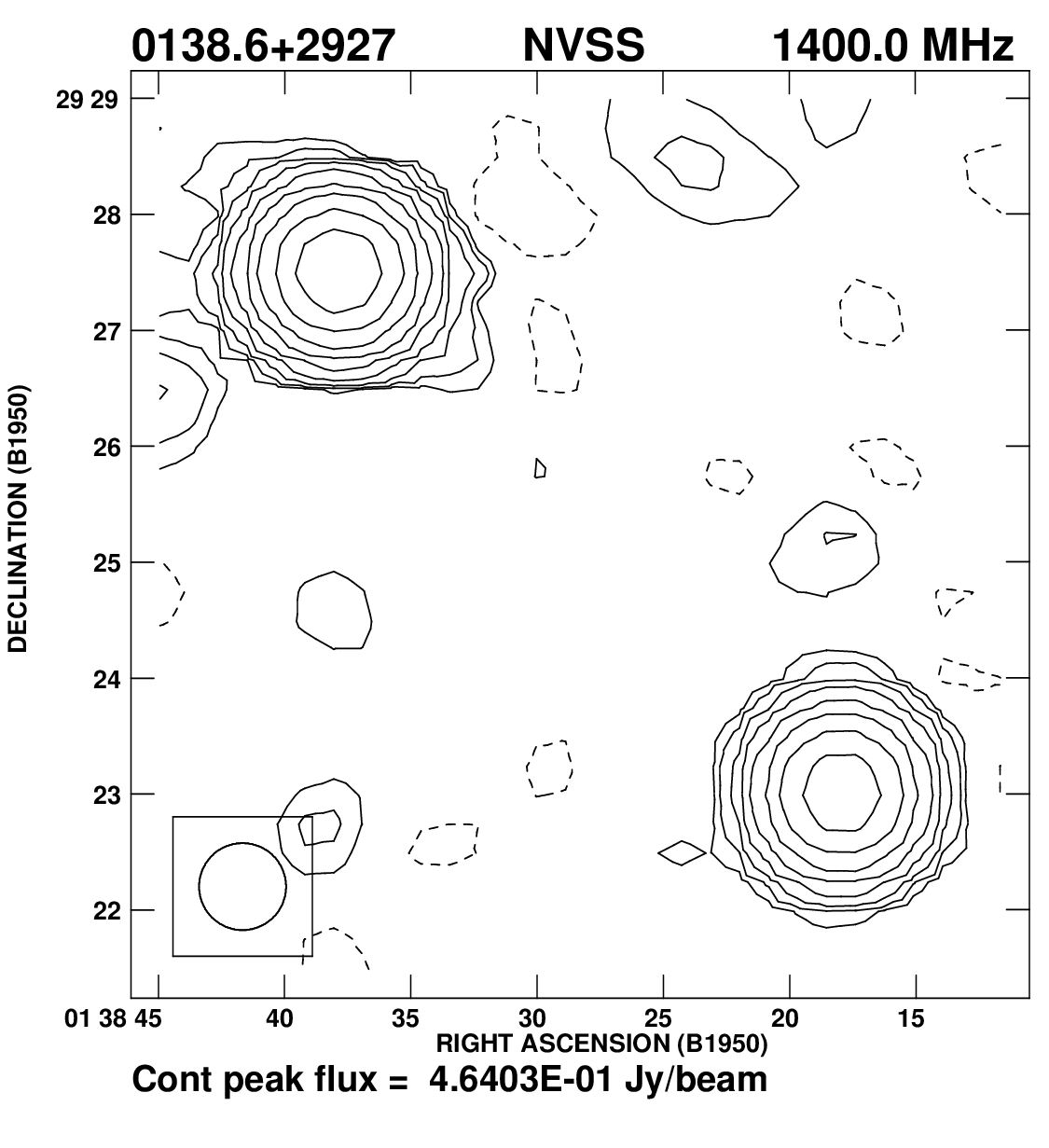,width=1.72in}
  \psfig{file=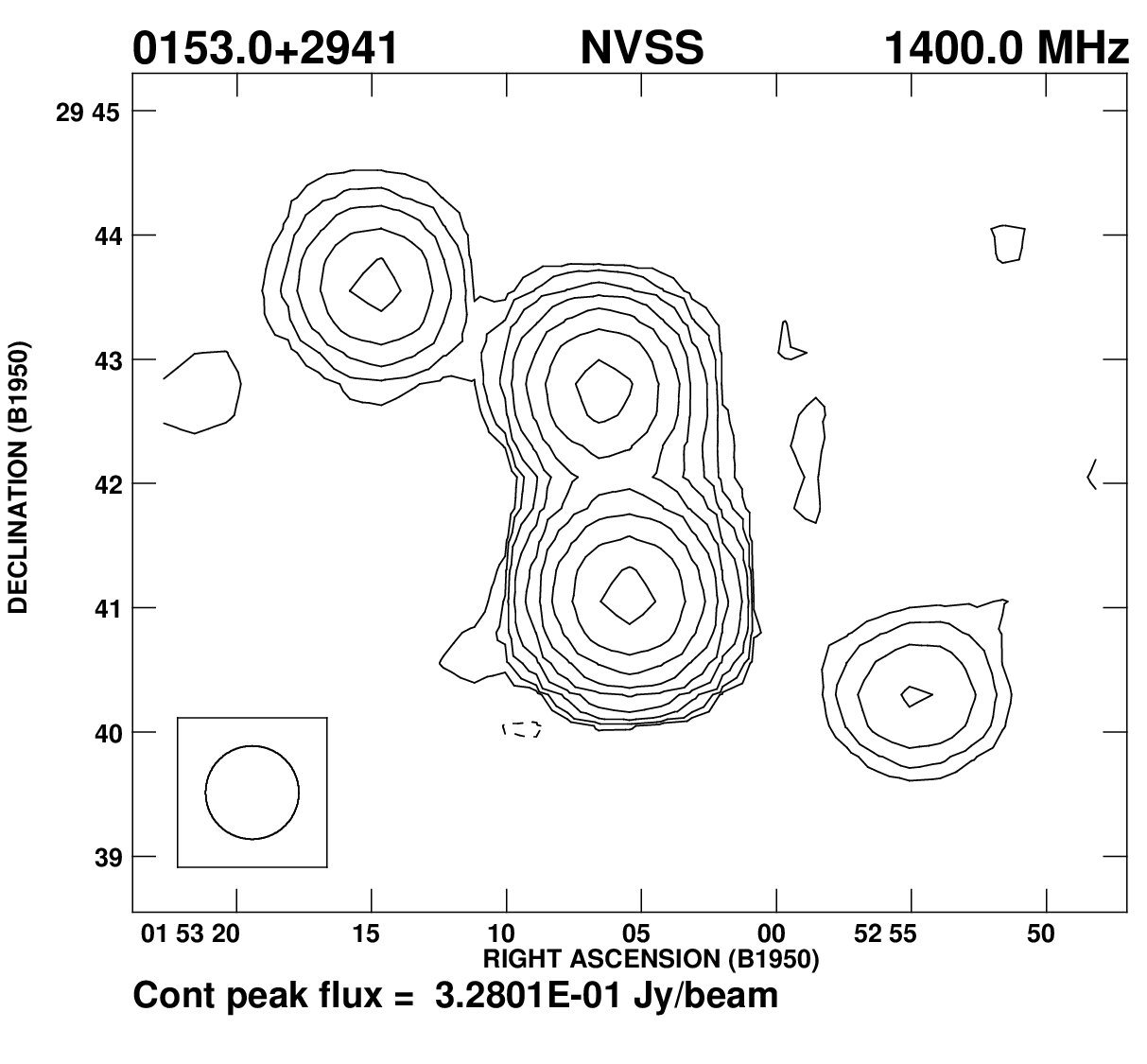,width=1.72in,angle=-90}
  \psfig{file=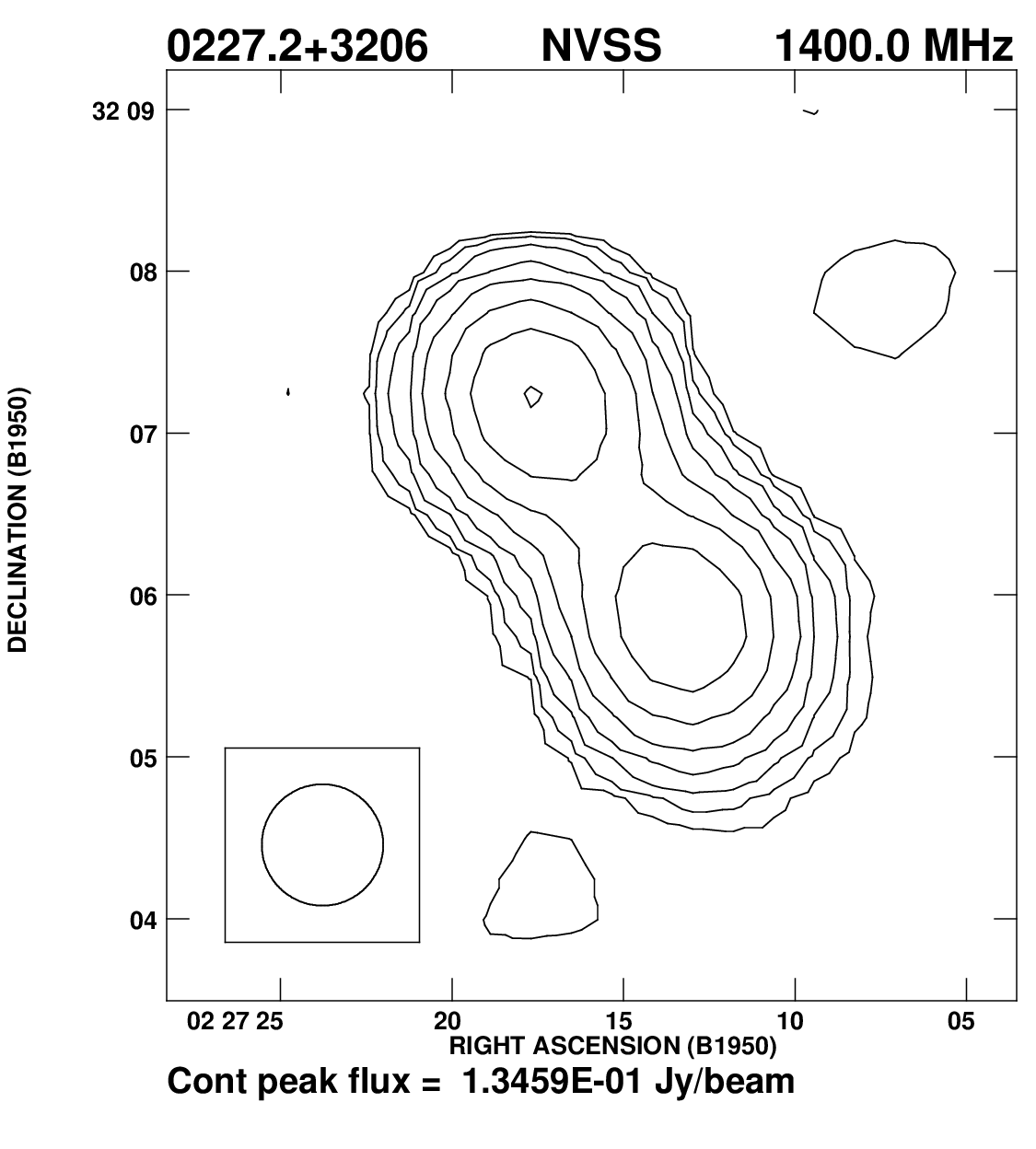,width=1.7in}
     }
}
\caption[]{NVSS images of the sources discussed in the text. The contour 
levels for all the images are -1, 1, 2, 4, 8, 16, \ldots mJy/beam. The
peak brightness in units of Jy/beam is given below each image.
}
\end{figure*}

\section{Observations, analysis and results}
Eighteen steep-spectrum sources ($\alpha_{408}^{5000}\gapp$0.5 (P81), where 
S$\propto\nu^{-\alpha}$)
from the B2 sample of intermediate strength were 
observed in snapshot mode with the Very Large Array (VLA) A-array at 4835 MHz on
1985 February 10. All of these were either unresolved or only 
slightly resolved by the WSRT 5-GHz observations of P81 made with an angular 
resolution of about 6 $\times$ 6cosec$\delta$ arcsec$^2$, where $\delta$ is the
declination of the source. 
In addition to these 18, the sample source 0204.6+3137 which was listed by
P81 as having a spectral index $\alpha_{408}^{5000}$=0.47 was also observed. 

The data were calibrated at the VLA and reduced using standard imaging and
CLEANing procedures in the AIPS (Astronomical Image Processing
System) package. All flux densities are on the Baars et al. (1977) scale.

The VLA radio images for the 18 resolved objects are presented in
Figs. 1 and 2. The observational parameters, and some of the properties
derived from the images, are listed for all 19 sources in Table 1,
which is arranged as follows.
Col. 1: source name in the IAU format using the B1950 co-ordinates; 
col. 2: optical classification for 
those objects with confirmed identifications, where G denotes a galaxy and Q a quasar;
col. 3: redshift; col. 4: the major and minor axes of the restoring 
beam in arcsec and its
position angle (PA) in degrees; col. 5: the rms noise in the total-intensity 
images in units of $\mu$Jy/beam; col. 6: component designations for the structures
revealed by the VLA images, with RS denoting the entire radio source for 
those without well-defined double-lobed structure; 
cols. 7 and 8: the peak and total flux density of the components
in units of mJy/beam and mJy; col. 9: the total flux density of the source 
in units of mJy determined by integrating over a box 
containing all emission detected in the VLA images;
col. 10: the classification of the radio structure of the source as determined 
from our images or those available in the literature. Here D denotes a double,
T a triple, CJ a core-jet source, HT a head-tail source, C+E a core with some 
extension, while SR is a slightly resolved source and U an unresolved source. 
Col. 11: the Fanaroff-Riley
classification for double-lobed and triple sources based on an
examination of the radio image.
A question mark (?) in cols. 10 or 11 denotes an uncertainty in the classification. 
Col. 12: the largest angular size (LAS) of the source in arcsec. These are usually
measured from the outermost peaks of radio emission. Col. 13: Here n denotes 
that there is a note on the source in Sect. 3.1 of this paper. 

For the 33 sample sources not observed by us, Table 2 lists the source
name, optical identification, redshift, structural classification,
Fanaroff-Riley class, largest angular size and references.  The
above parameters have the same meaning as in Table 1.  The sources
for which the LAS has been estimated from the NVSS (Condon et al. 1998) 
images have a superscript N in col. 6.

\subsection{Notes on individual sources}
Eight images from the NVSS survey (Condon et al. 1998) are presented in Fig. 3 as
these provide significant extra information on sample sources, both those observed
and not observed by us.

\noindent
{\bf 0003.0+3149} The NVSS image shows three non-collinear components (Fig. 3).
   It is not clear whether they are all related. Using the WENSS 
   (Rengelink et al. 1997) and NVSS (Condon et al. 1998) flux
   densities, the spectral index of the dominant component is 
   $\alpha_{327}^{1400}$ $\sim$0.40 while 
   that of the two weaker components in order of increasing right
   ascension are $\sim$0.87 and 0.82 respectively.

\noindent
{\bf 0008.7+3200} The position of the optical galaxy at RA(B1950) 00$^h$ 08$^m$
   43.$^s$6 and Dec(B1950) 32$^\circ$ 00$^\prime$ 33$^{\prime\prime}$ (P81) lies 
   between the two extended lobes of radio emission seen in the NVSS image (Fig. 3). 
   The largest angular size measured from the edges of the radio source is close
   to about 500 arcsec, yielding a projected linear size of $\sim$1.3 Mpc in a
   Universe with q$_o$=0.5 and H$_o$=50 km s$^{-1}$ Mpc$^{-1}$. 
   This makes it a candidate giant radio galaxy (cf. Ishwara-Chandra \& Saikia
   1999) with a somewhat unusual non-collinear structure. A higher resolution image
   would be useful to clarify  the structure.

\noindent
{\bf 0039.0+3208} The position of the optical galaxy at RA(B1950) 00$^h$ 39$^m$
   04.$^s$94 and Dec(B1950) 32$^\circ$ 08$^\prime$ 27$^{\prime\prime}$.3 
   (P81) is close to that of the brighter eastern lobe in the
   NVSS image (Fig. 3). The western and eastern radio components on opposite sides
   of the optical galaxy seen in the WSRT image by P81 are separated from the 
   optical galaxy by $\sim$17.4 and 14.4 arcsec respectively. The more extended
   emission on both sides seen in the NVSS image suggests that this is a candidate
   double-double radio galaxy (Schoenmakers et al. 2000) and merits further investigation.

\noindent
{\bf 0051.1+3058} The angular size has been estimated from the NVSS image (Fig. 3). 
    The weaker NVSS component to the south-west may be related, but 
    appears to be resolved out in the image by P81.

\noindent
{\bf 0110.0+3152} The LAS given in Table 2 is between the prominent peaks seen in the image
    made by Wilkinson et al. (1998). A more extended halo is visible in a lower-
    resolution image (Hutchings et al. 1998).   

\noindent
{\bf 0125.1+3229} The image resembles a single lobe in a double-lobed source. Two additional,
    weaker radio sources are visible to the south and east in the NVSS image (Fig. 3), but
    it is not clear if either of these is related to the B2 source.

\noindent
{\bf 0138.6+2927} The NVSS image (Fig. 3) shows a prominent component to the south-west.
    It is not clear if it is related to the source. The spectral indices $\alpha$(327-1400
    MHz) are $\sim$0.80 and 0.85 for the source and the south-western component respectively.

\noindent
{\bf 0153.0+2941} The LAS given by P81 is listed. The NVSS image (Fig. 3) shows four 
    components roughly symmetric about RA(B1950) 01$^h$ 53$^m$
    05$^s$ and Dec(B1950) 29$^\circ$ 42$^\prime$. A sensitive higher resolution
    image should help clarify any relationship between the components.

\noindent
{\bf 0204.6+3137} The VLBA image shows a core-jet structure oriented along a PA$\sim$
    135$^\circ$ (Peck \& Taylor 2000). Our image suggests a strongly curved jet with 
    the most distant component being along a PA$\sim$$-$90$^\circ$.

\noindent
{\bf 0227.2+3206} P81 note that the source is resolved out in the WSRT observations. 
    The source appears double-lobed in the NVSS image (Fig. 3), and the LAS has been 
    estimated from this.

\noindent
{\bf 2330.5+2937} The source appears unresolved in our observations, although its spectral
    index, $\alpha_{408}^{5000}$ $\sim$0.95 (P81). 

\noindent
{\bf 2342.5+2926} The spectral index between 327 MHz and 5 GHz is about 0.7. It has a highly
    asymmetric radio structure, and has been detected in the 2MASXi survey (NED).

%
%

\begin{table}
{\bf Table 2.} Properties of other sources in the sample \\

\begin{tabular}{l l l l l r l}
\hline
Source & O    & z & Str. & FR & LAS & Refs. \\
       &      &   &      &    & $^{\prime\prime}$ & \\
\hline
0003.0+3149 &    &        &    &    &         &n  \\
0007.5+3028 &    &        & D? &    & 57$^N$  &   \\
0007.8+3312 & Q  & 0.743  & T  & II & 78      &1,2\\
0008.7+3200 & G  & 0.1073 & D  & II?&357$^N$  &3,4,n\\
0013.5+3222 &    &        & D  & II & 54      &4  \\
0014.4+3152 & Q  & 1.086  & D  & II & 17      &4  \\
0015.9+3049 &    &        & D  & II & 49      &4  \\
0039.0+3208 & G  &        & D  & II &106$^N$  &n  \\
0039.9+2949 &    &        & D  & II & 12      &4  \\
0051.1+3058 &    &        & D  & II?& 85$^N$  &n  \\
0055.0+3004 & G  & 0.0167 & D  & II?&3480     &5,6\\
0110.0+3152 & Q  & 0.6030 & CJ?&    &0.7      &7,8,n\\
0110.6+2942 & Q  & 0.363  & T  & II & 76      &9  \\
0138.6+2927 &    &        &    &    &         &n  \\
0149.6+3335 & Q  & 2.4310 & C+E&    & 4       &7,10\\
0153.0+3001 &    &        & D  &    & 14      &4  \\
0153.0+2941 &    &        & D  &    & 13      &4,n\\
0154.3+3159 & G  & 0.0894 & HT & I  & 102     &11 \\
0157.8+3323 & Q  &        &  T & II & 25      &4,12\\
0200.8+3027 &    &        & C+E& II & 69      &4  \\
0202.1+3158 & Q  & 1.466  & CJ &    &0.04     &13,14\\
0206.6+3340 &    &        & SR &    & 9       &16 \\
0216.8+3334 &    &        &  D & II & 12      &17 \\
0217.8+3227 & Q  & 1.620  & CJ?&    & 3       &7,18  \\
0222.3+3105 & Q  & 2.982  & CJ?&    &         &18 \\
0226.8+2924 &    &        &  D & II & 34      &4  \\
0227.2+3206 &    &        &  D & II?& 97$^N$  &n  \\
2330.5+3038 &    &        & SR &    &$\sim$5  &4,15 \\
2335.0+3007 &    &        &  D & II?& 39      &4  \\
2341.8+3019 &    &        &  D & II?& 19      &4  \\
2347.8+3013 & G  & 0.374  &  D & II?& 37      &4,19,20,21 \\
2348.5+3003 &    &        &  T?& II & 99      &4  \\
2349.8+3247 & Q  & 0.659  &  T & II & 60      &1,22,23\\
\hline
\end{tabular}
References:
1: Price et al. 1993; 2: Wardle \& Potash 1985; 3: NED;
4: Padrielli et al. 1981; 5: Bridle et al. 1976;
6: Willis et al. 1981; 7: Wilkinson et al. 1998;
8: Hutchings et al. 1998; 9: Garrington et al. 1991; 
10: Neff \& Hutchings 1990; 11: Owen \& Ledlow 1997;
12: Minter \& Spangler 1996; 13: Fey \& Charlot 2000;
14: Fomalont et al. 2000; 15: Rudnick \& Adams 1979;
16: Fanti \& Padrielli 1977; 17: R\"{o}ttgering et al. 1994;
18: Willott et al. 1998; 19: Zirbel 1997; 20: Nilsson et al. 1993; 
21: Grueff et al. 1981; 22: Potash \& Wardle 1980; 
23: Bogers et al. 1994 
\end{table}


\begin{figure}
\vbox{
  \psfig{file=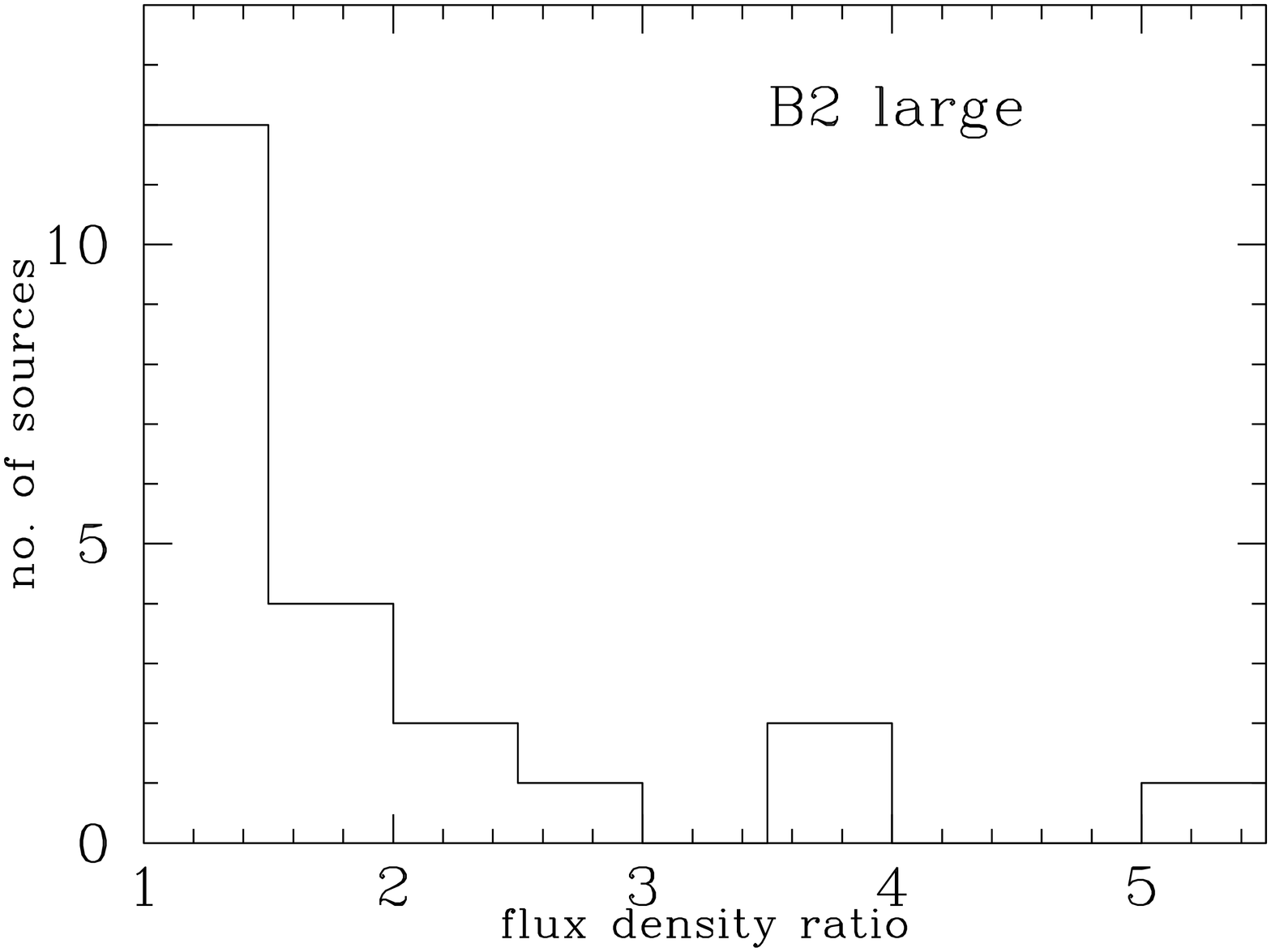,width=3.0in,angle=-90}
  \psfig{file=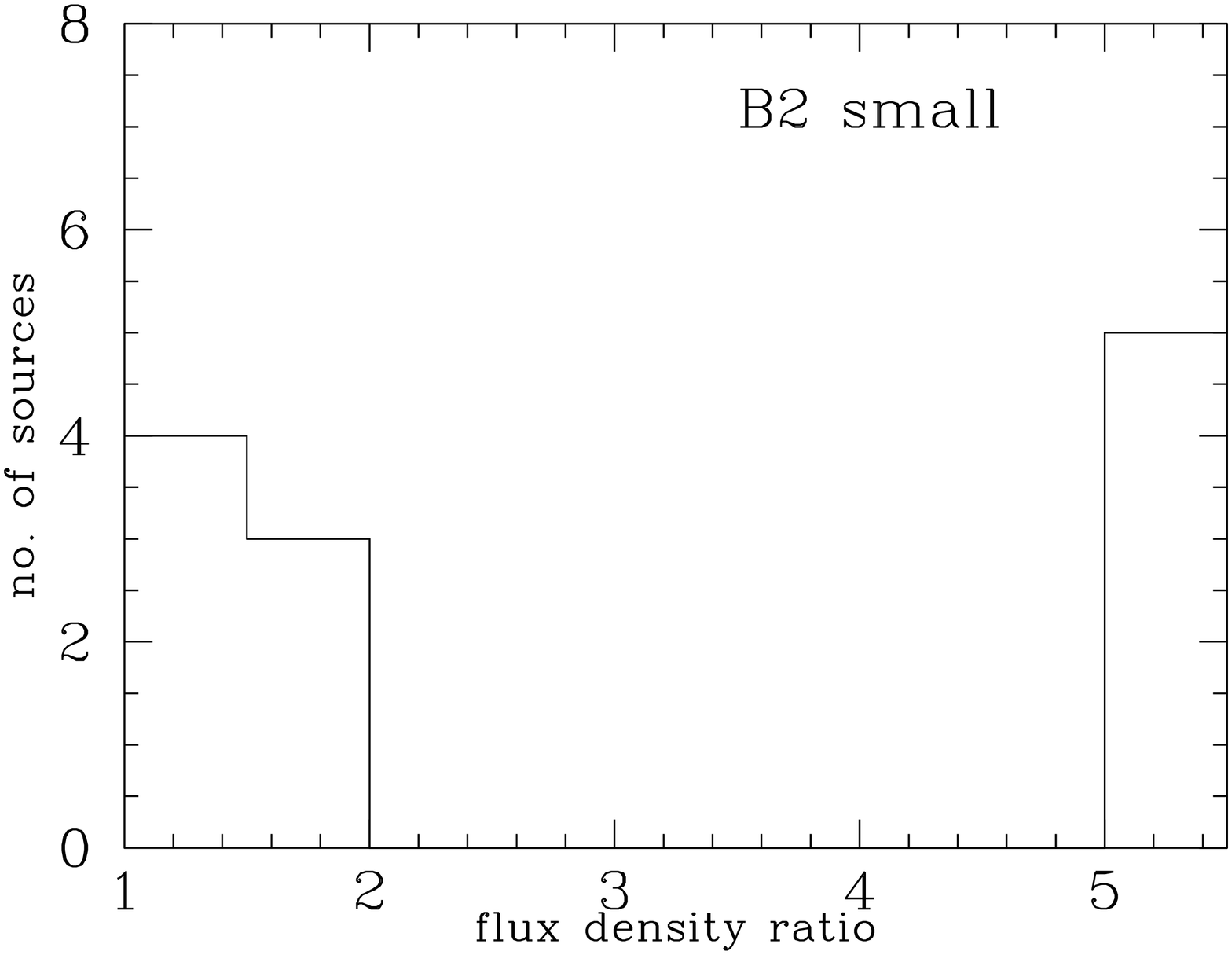,width=3.0in,angle=-90}
   }
\caption[]{The distributions of the flux density
ratio, R$_s$ for the B2 sources discussed in the
text. All the sources with R$_s$ $>$5 have been
placed in the last bin. 
}
\end{figure}

\begin{figure}
\vbox{
  \psfig{file=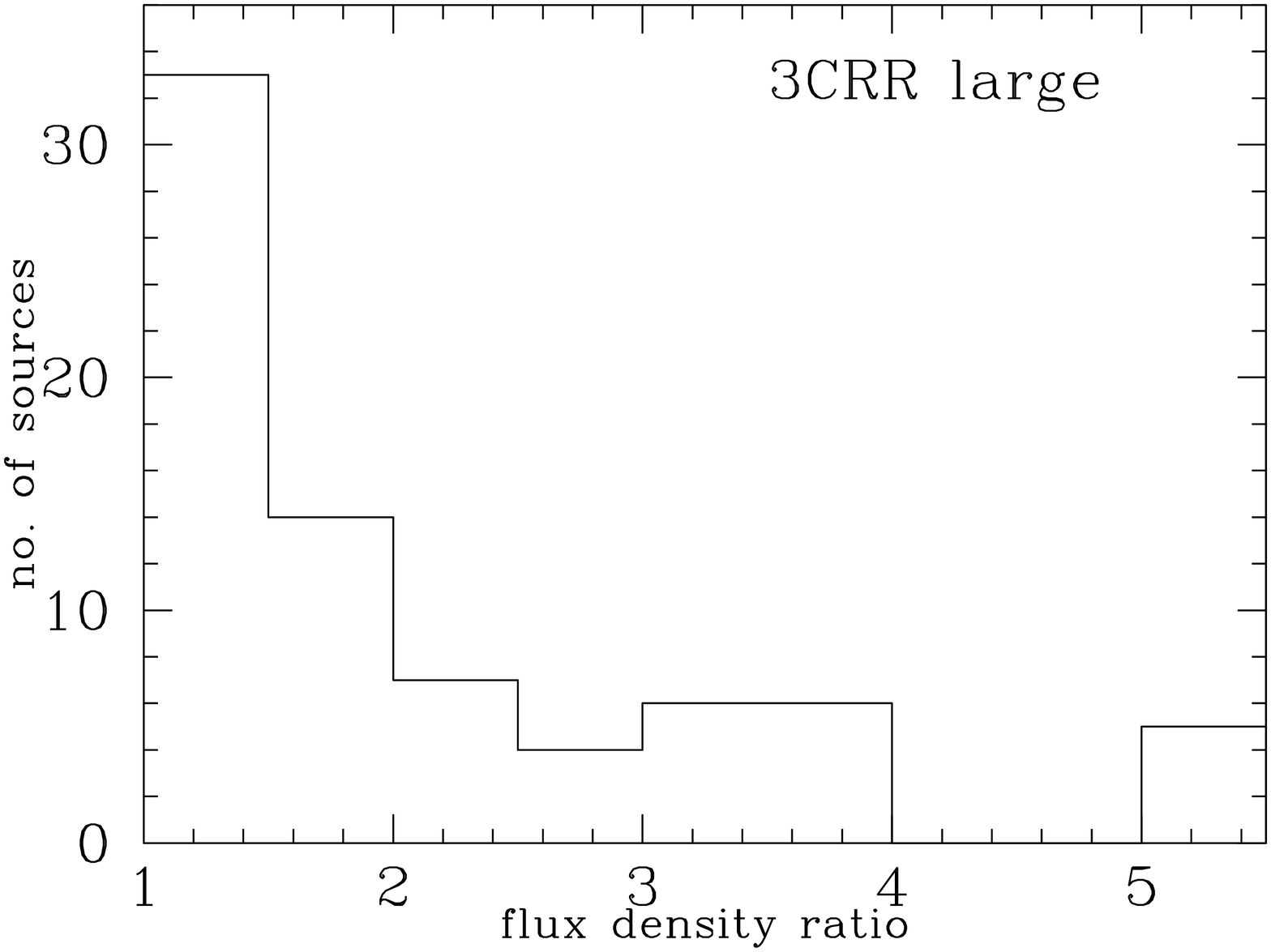,width=3.0in,angle=-90}
  \psfig{file=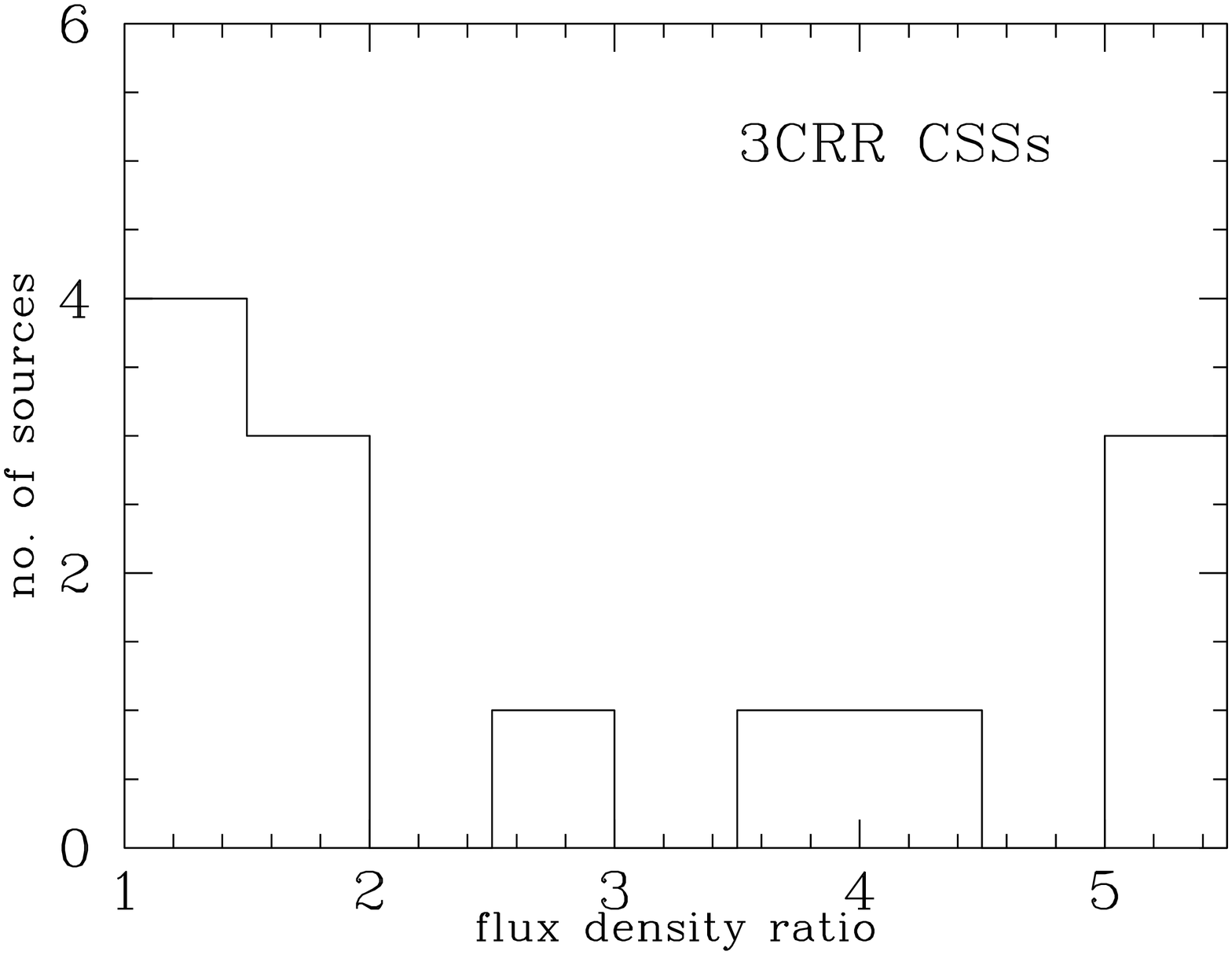,width=3.0in,angle=-90}
   }
\caption[]{The distributions of the flux density
ratio, R$_s$ for the 3CRR sources discussed in the
text. All the sources with R$_s$ $>$5 have been
placed in the last bin. 
}
\end{figure}

\section{Discussion}

Of the 19 sources observed by us, 13 have well-defined, double-lobed structures of
which two also have detected central components. Of the remaining 6, two appear
possibly double-lobed, two are single sources, one (0204.6+3137) has a 
somewhat flatter radio spectrum 
and appears to have a core-jet structure, while 0125.1+3229 has a complex
structure. The double-lobed radio sources tend to have Fanaroff-Riley class II
structures with hot-spots at the outer edges of the sources. Unfortunately, the optical
identification and redshift information for this sample is still very incomplete 
for investigating in detail the source properties and all symmetry parameters. Further,
radio cores have been detected in only a small fraction of the sources. Therefore,
we examine only the flux density ratio, R$_s$, of opposing lobes of the
steep-spectrum double-lobed sources, classified as D or T in Tables 1 and 2. This 
parameter can be used to  
probe asymmetries in the gas distribution in the central regions of these galaxies,
which might be related to the infall of gas which fuels the radio source.
The radio jet propagating outwards on the denser side will be more dissipative, and
hence should be brighter, and closer to the nucleus (cf. Eilek \& Shore 1989; Gopal-krishna 
\& Wiita 1991).  Although relativistic motion of the lobes could also
affect their brightness asymmetry, this alone is not likely to generate highly asymmetric 
sources since the velocity of advancement is expected to be only mildly relativistic, and
most of our objects are likely to be inclined at large angles to the line of sight. 
Also, while the hotspots advance outwards,  the rest of the radio lobe 
emission is due to backflow from these high-brightness features. For a velocity of 
advancement of $\sim$0.1c (cf. Scheuer 1995), the brightness asymmetry of the lobes for
a source inclined at 30$^\circ$ to the line of sight is $\lapp$2. Nevertheless, for a 
few quasars, the brightness asymmetry may be enhanced due to relativistic motion of the 
hotspots. However, this is not likely to be an important factor for the  objects in our sample 
because in addition to the low velocities, these are lobe-dominated objects 
and are unlikely to be inclined 
at small angles to the line of sight. Also, most of the objects ($\sim$75 per cent) are not 
identified with quasars, and are therefore expected to be inclined
at $>$45$^\circ$ to the line of sight, i.e. the dividing line between radio galaxies and
quasars in the unified scheme (Barthel 1989), thereby further decreasing the expected
asymmetry due to the velocity of advancement.

Considering the steep-spectrum, double-lobed B2 sources, we define the `small-source' 
sample to consist of those with an LAS less than 10 arcsec. 
The remaining objects constitute the `large-source' sample. The median angular size of the
sample of small sources is about 4 arcsec compared with about 40 arcsec for the larger
sources. The ratio of flux density of the oppositely directed lobes, R$_s$, 
has been estimated for all these sources from either our 
observations or the references listed in Table 2.  R$_s$ ranges
from $\sim$1.4 to 20 for the small-source sample, and has a median value of about 2.
For the large-source sample, R$_s$ ranges from $\sim$1.0 to 7,   
with a median value of $\sim$1.4 (Fig. 4). 

It can be seen from Fig. 4 that a significant fraction of the small sources show large
asymmetries. Defining a very asymmetric source to be one with R$_s$ $>$ 5, 
we find that 5 of the 12 small-sized objects ($\sim$40 per cent) are highly asymmetric,
while only one of the 22 ($\sim$5 per cent) objects in the large-source B2 sample 
have R$_s$ $>$ 5. It is of interest to compare this with earlier studies of 
the flux density ratio for sources in the well-known 3CRR sample
(e.g. Saikia et al. 1995, 2001; Arshakian \& Longair 2000). 
For example, for the 3CRR sources studied by Saikia et al., which was confined 
to the high-luminosity FRII sources, the median values of R$_s$ for the large sources
and the CSSs are about 1.8 and 2 respectively. This is similar to that of the B2 sources.  
The distribution for the CSSs again appears to have a number of very asymmetric
sources with 3 of the 13 sources (about 23 per cent) having a value of R$_s$$>$5,
while only 5 of the 75 ($\sim$7 per cent) non-CSS 3CR sources exhibit such a high
degree of asymmetry (Fig. 5). 
The small-sized sources in both the B2 and 3CRR samples appear 
to have a greater fraction of objects exhibiting high flux-density asymmetry 
compared with the larger sources. Excluding these highly asymmetric CSS objects,
the remainder of the population of CSS sources show very similar brightness ratios
to those of the larger objects. 

\begin{figure}
\vbox{
  \psfig{file=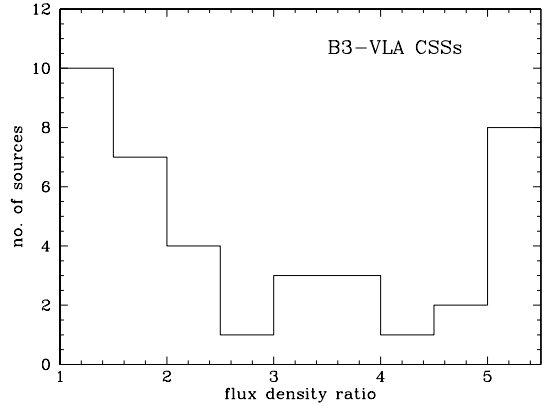,width=3.0in,angle=-90}
   }
\caption[]{The distributions of the flux density
ratio, R$_s$ for the B3-VLA CSS sources discussed in the
text. All the sources with R$_s$ $>$5 have been
placed in the last bin. 
}
\end{figure}

We have also compared this result for the CSS objects with the large sample of CSSs
selected from the B3-VLA sample with S(408) $\geq$ 0.8 Jy (Fanti et al. 2001). 
Fanti et al. have classified all sources with two distinct steep-spectrum
($\alpha\gapp$0.5) components as double, and those with three distinct steep-spectrum
components as triple. We consider all the doubles, plus just one triple source, 0744+464.
This source is collinear and the central feature, which has a flatter spectrum, 
contributes less than a few per cent of the total flux density and does not 
significantly affect the flux density ratio.  The R$_s$ distribution for the
39 sources from this sample is shown in Fig. 6. The median value of 
R$_s$ for the B3-VLA CSS sample is about 2.4, with 8 of the 39 objects ($\sim$21 per cent)
having a value of R$_s$ $>$ 5. This is consistent with the results for
the B2 and 3CRR samples.

To assess the statistical significance of the results, we have considered all the
CSS sources from the B2, 3CRR and B3-VLA samples together and compared their R$_s$
distribution with that of the larger sources from the B2 and 3CRR samples. This
increases the sample sizes to a total of 64 CSS objects and 97 larger sources, and
should yield a statistically reliable result. Considering the above samples, the 
median values of R$_s$ for the CSS objects and the larger sources are about 2 and 1.5
respectively. A Kolmogorov-Smirnov test shows the two distributions to be different 
at a confidence level of greater than 99 per cent. However, if we exclude the most
asymmetric objects, i.e. those with R$_s$ $>$5, the distributions for the rest of the
sources are not significantly different. The fraction of such asymmetric sources
for the combined sample is 25 per cent for the CSSs and 6 per cent
for the larger sources. If the threshold for defining a very asymmetric source is
changed to R$_s$ $>$4, the fraction for the CSS objects increases to about 30 per cent,
while that for the larger objects remains the same. These trends are also seen if the
confirmed quasars and the remaining objects are considered separately.

Thus, in about 25 per cent of the CSSs the flux density asymmetry is
significantly increased, possibly due to interaction of one of the jets with a much 
denser external medium or a dense cloud on one side of the nucleus. These density
asymmetries might be intimately related to the infall of gas which fuels the radio source. 
Taking the typical distances of these clouds of gas from the nucleus of the galaxy to 
be about 5 to 10 kpc, the fraction of very asymmetric sources found above suggests
that the size of the ``clouds'' should be about 3 to 7 kpc. This is typical of the 
sizes of dwarf galaxies (cf. Swaters 1999). 
 
It is also of interest to enquire whether the brightness asymmetries of 
CSSs depend on cosmic epoch because of the larger incidence of interactions and
mergers in the past, as seen in Hubble Space Telescope studies of distant galaxies (cf.
Abraham et al. 1996, Brinchmann et al. 1998, Ellis et al. 2000). For this, we have
considered the double-lobed CSSs with a measured or estimated redshift 
from the combined sample. The R$_s$-redshift diagram
for these 43 sources is presented in Fig. 7. It is quite clear from 
the figure that the flux density asymmetry shows no significant dependence on redshift.
This suggests that although interactions and  mergers may increase with redshift globally,
the density asymmetries in these CSS objects which are young and possibly still being
fuelled by the infall of gas, are similar at different redshifts. 

\begin{figure}
\vbox{
  \psfig{file=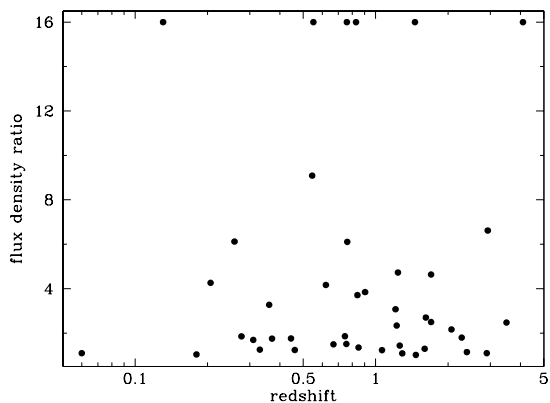,width=3.0in,angle=-90}
   }
\caption[]{The R$_s$$-$redshift diagram for the CSSs in the 3CRR, B2
and B3-VLA samples discussed in the text. The six sources at the upper part of the
figure have R$_s$ $>$16.
}
\end{figure}

\section{Conclusions}
Radio images of small sources from the B2 sample of intermediate strength have been 
presented. About 25 per cent of the CSSs from the B2, 3CRR and the B3-VLA samples
exhibit a higher degree of brightness asymmetry, with the flux density ratio
R$_s$ $\gapp$5.  
The fraction of such highly asymmetric objects amongst the larger sources is only 
about 6 per cent. The larger fraction of highly asymmetric objects amongst CSS sources
might be due to interactions of the radio plasma with infalling clouds of gas which fuel
the radio source. The size of such a cloud is estimated to be about a few kpc, which is
typical of dwarf galaxies. The brightness asymmetry for these objects shows no significant
dependence on redshift, although such a situation might be expected due to the 
larger incidence of interactions and mergers in 
the past. This suggests that the density asymmetries in these young radio sources are
similar at different redshifts.

\begin{acknowledgements}
We thank the referee, Ignas Snellen, for several constructive comments which 
considerably improved the paper, and also for 
reminding us of the paper by Carla Fanti and her collaborators on the B3-VLA sample.
The National Radio Astronomy Observatory is a facility of the National Science Foundation
operated under co-operative agreement by Associated Universities, Inc.  We have made use of the
NASA/IPAC Extragalactic Database (NED), which is operated by the Jet Propulsion Laboratory,
California Institute of Technology under contract with the National Aeronautics and Space
Administration. We thank Dave Rossi for preserving the 9-inch tape containing the
calibrated data in his loft for many years, and Dave Shone for managing to read the tape!
One of us (DJS) would like to thank the PPARC Visitors Programme at Jodrell Bank Observatory
and Ralph Spencer who looks after this programme, for financial support, and Peter Thomasson 
for hospitality while most of this work was done.
\end{acknowledgements}

\end{document}